\PassOptionsToPackage{dvipsnames}{xcolor}
\documentclass[twocolumn]{aastex631}

\usepackage{amsmath}
\usepackage{float}

\begin{document}

\title[Massive star winds]{Implications of modern mass-loss rates for massive stars}

\author{JD Merritt}
\affiliation{Institute for Fundamental Science, Department of Physics, University of Oregon, Eugene, OR 97403, USA}

\author[0000-0002-6100-537X]{Simon Stevenson}
\affiliation{Centre for Astrophysics and Supercomputing, Swinburne University of Technology, Hawthorn, VIC 3122, Australia}
\affiliation{ARC Center of Excellence for Gravitational-wave Discovery (OzGrav), Melbourne, Australia}

\author[0000-0002-2090-9751]{Andreas Sander}
\affiliation{Zentrum f{\"u}r Astronomie der Universit{\"a}t Heidelberg, Astronomisches Rechen-Institut, M{\"o}nchhofstr. 12-14, 69120 Heidelberg, Germany}
\affiliation{Interdisziplin{\"a}res Zentrum f{\"u}r Wissenschaftliches Rechnen, Universit{\"a}t Heidelberg, Im Neuenheimer Feld 225, 69120 Heidelberg, Germany}

\author[0000-0002-6134-8946]{Ilya Mandel}
\affiliation{School of Physics and Astronomy, Monash University, Clayton VIC 3800, Australia}
\affiliation{ARC Center of Excellence for Gravitational-wave Discovery (OzGrav), Melbourne, Australia}

\author{Jeff Riley}
\affiliation{School of Physics and Astronomy, Monash University, Clayton VIC 3800, Australia}
\affiliation{ARC Center of Excellence for Gravitational-wave Discovery (OzGrav), Melbourne, Australia}

\author[0000-0002-2916-9200]{Ben Farr}
\affiliation{Institute for Fundamental Science, Department of Physics, University of Oregon, Eugene, OR 97403, USA}
\affiliation{Centre for Astrophysics and Supercomputing, Swinburne University of Technology, Hawthorn, VIC 3122, Australia}
\affiliation{School of Physics and Astronomy, Monash University, Clayton VIC 3800, Australia}
\affiliation{ARC Center of Excellence for Gravitational-wave Discovery (OzGrav), Melbourne, Australia}

\author[0000-0001-5484-4987]{L.~A.~C.~van~Son}
\affiliation{Center for Computational Astrophysics, Flatiron Institute, 162 Fifth Avenue, New York, NY 10010, USA}
\affiliation{Department of Astrophysical Sciences, Princeton University, 4 Ivy Lane, Princeton, NJ 08544, USA}

\author[0000-0001-6147-5761]{Tom Wagg}
\affiliation{Department of Astronomy, University of Washington, Seattle, WA, 98195}

\author{Serena Vinciguerra}
\affiliation{Anton Pannekoek Institute for Astronomy, University of Amsterdam, Netherlands}

\author{Holden Jose}
\affiliation{Institute for Fundamental Science, Department of Physics, University of Oregon, Eugene, OR 97403, USA}

\begin{abstract}
Massive stars lose a significant fraction of their mass through stellar winds at various stages of their lives, including on the main sequence, during the red supergiant phase, and as helium-rich stripped stars.
In stellar population synthesis, uncertainty in the mass-loss rates in these evolutionary stages limits our understanding of the formation of black holes and merging compact binaries. 
In the last decade, the theoretical predictions, simulation, and direct observation of wind mass-loss rates in massive stars have improved significantly, typically leading to a reduction in the predicted mass-loss rates of massive stars. 
In this paper we explore the astrophysical implications of an updated treatment of winds in the COMPAS population synthesis code. 
There is a large amount of variation in predicted mass-loss rates for massive red supergiants; some of the prescriptions we implement predict that massive red supergiants are able to lose their hydrogen envelopes through winds alone (providing a possible solution to the so-called missing red supergiant problem), while others predict much lower mass-loss rates that would not strip the hydrogen envelope.
We discuss the formation of the most massive stellar-mass black holes in the Galaxy, including the high-mass X-ray binary Cygnus X-1 and the newly discovered Gaia BH3. 
We find that formation rates of merging binary black holes are sensitive to the mass-loss rate prescriptions, while the formation rates of merging binary neutron stars and neutron-star black hole binaries are more robust to this uncertainty. 

\end{abstract}


\section{Introduction} 
\label{sec:intro}

Massive stars (with initial masses $> 8$\,M$_\odot$) experience significant mass loss through stellar winds throughout their lives, losing a sizable fraction, and sometimes nearly all, of their initial mass.
These stars are the progenitors of supernovae, neutron stars and black holes.
There are large uncertainties in both theoretical predictions for, and observations of, the mass-loss rates of massive stars.

Massive stars preferentially form in binaries \citep[e.g.,][]{Sana:2012Sci}.
The evolution of massive binaries can lead to the formation of merging compact binaries \citep[e.g.,][]{1973NInfo..27...70T, 1973A&A....25..387V, 2021hgwa.bookE..16M, 2022PhR...955....1M}.

The landscape of compact binary astrophysics is expanding quickly with the rapidly growing catalogs of compact binary mergers detected in gravitational waves (GWs) by the LIGO-Virgo-KAGRA (LVK) collaboration \citep[][]{Abbott:2023PhRvXGWTC3,Abac:2025GWTC4}, and the detection of detached Galactic black hole binaries by Gaia \citep[][]{GaiaBH2El-Badry, El-Badry:2024GaiaBH3, GaiaBH32024}. 
One of the greatest uncertainties in determining the evolutionary fates of massive stars and binaries is mass loss through stellar winds.

In recent years, increasingly sophisticated models of mass loss have been developed for various phases of stellar evolution \citep[e.g.,][]{Krticka:2017aap,Krticka:2018aap,Sander:2020MNRAS}.
At the same time, empirical mass loss prescriptions for evolved stars have benefited from increasingly large samples of massive stars from new observational surveys \citep[e.g.,][]{Yang:2023arXiv,Shenar:2019aap}. 

When incorporated into detailed stellar evolution or stellar population synthesis codes, these updated wind prescriptions allow us to improve the modeling of a plethora of products of massive binary stellar evolution.
We can generate populations of compact binary mergers and compare them to recent GW observations \citep[e.g.,][]{Broekgaarden:2022MNRAS,Stevenson:2022MNRAS}, populations of massive supernova progenitors observed in the local Universe \citep[e.g.,][]{Smartt_2009,Davies_2020} and stellar-mass black holes in binaries in the Milky Way \citep[e.g.,][]{Miller-Jones:2021plh,GaiaBH32024}.

In this paper we update the prescriptions for wind mass loss of massive stars in the rapid binary population synthesis code COMPAS \citep[][]{COMPASTeam:2021tbl} during several phases of stellar evolution, including on the main sequence (MS) for both massive and very massive stars (VMS), during the red supergiant (RSG) phase and for helium-rich stripped stars (which we refer to as Wolf-Rayet or WR stars). 
VMS and RSG are new additions to the phase-based treatment of mass loss in COMPAS. In total, we add 15 new wind-mass-loss prescriptions. 
We find that uncertainties in mass-loss prescriptions have the largest impact on predictions for the formation rate of merging binary black holes, while the formation rates of merging binary neutron stars and neutron star-black hole binaries are more robust.
We also discuss our results in the context of the missing red supergiant problem \citep[][]{Smartt_2009}, the maximum black hole mass at solar metallicity \citep[][]{Belczynski:2010ApJ,Miller-Jones:2021plh, Neijssel:2021imj, Bavera:2023NatAs, Romagnolo:2024ApJL} and the formation of Gaia BH3 \citep[][]{GaiaBH32024, El-Badry:2024GaiaBH3}. 

The remainder of this paper is structured as follows: in Section~\ref{sec:methods} we introduce the population synthesis code COMPAS, and describe the mass-loss prescriptions we have implemented for various stages of massive stellar evolution.
In Section~\ref{sec:results}, we illustrate the impact of these updated mass-loss prescriptions on predictions for the remnants of massive stars, including the maximum black hole mass and populations of double compact objects.
We discuss our results and conclude in Section~\ref{sec:discussion}.

\begin{table*}[t]
\resizebox{\textwidth}{!}{\begin{tabular}{@{\extracolsep{\fill}} c | c c c c }
Prescription Set Name & OB & RSG & VMS & stripped He star \\ [0.5ex] 
\hline\hline
New Defaults ($\textsc{Merritt2025}$) & \citet{Vink:2021MNRAS} & \citet{Decin:2024AA} & \citet{Sabhahit:2023lni} & \citet{Sander:2023aap}, \citet{Vink:2017aap} \\
\hline
$\textsc{Pessimistic}$ & \citet{Vink:2021MNRAS} & \citet{VinkSabhahit:2023aap} & \citet{Bestenlehner:2020MNRAS} & \citet{Shenar:2019aap} \\
\hline
Old Defaults ($\textsc{Belczynski2010}$) & \citet{Vink:2000aap,Vink:2001aap} & \citet{Nieuwenhuijzen:1990aap} & \citet{Vink:2000aap,Vink:2001aap} & \citet{Hurley:2000MNRASSSE} \\
\hline
\end{tabular}}
\caption{Sets of mass-loss prescriptions used in this paper, where the columns are evolutionary phases.}
\label{table:prescriptions}
\end{table*}

\begin{table*}[t]
\begin{tabular}{@{\extracolsep{\fill}} c | c | c | c | c }
Parameter & OB & RSG & VMS & stripped He star \\ [0.5ex] 
\hline\hline
$T_\mathrm{eff}$(K) & $\geq$ 8000 & $<$ 8000 & -- & -- \\
\hline
$M$ (M$_{\odot}$)& $<$ 100 & $\geq$ 8 & $\geq$ 100 & -- \\
\hline
COMPAS Stellar Type & -- & Core helium burning or later & -- & Helium star \\
\hline
\end{tabular}\caption{Application criteria for mass loss based on evolutionary phase.}
\label{table:newtable}
\end{table*}

\section{Methods}
\label{sec:methods}

\begin{figure*}
    \centering
    \includegraphics[width=1.0\textwidth]{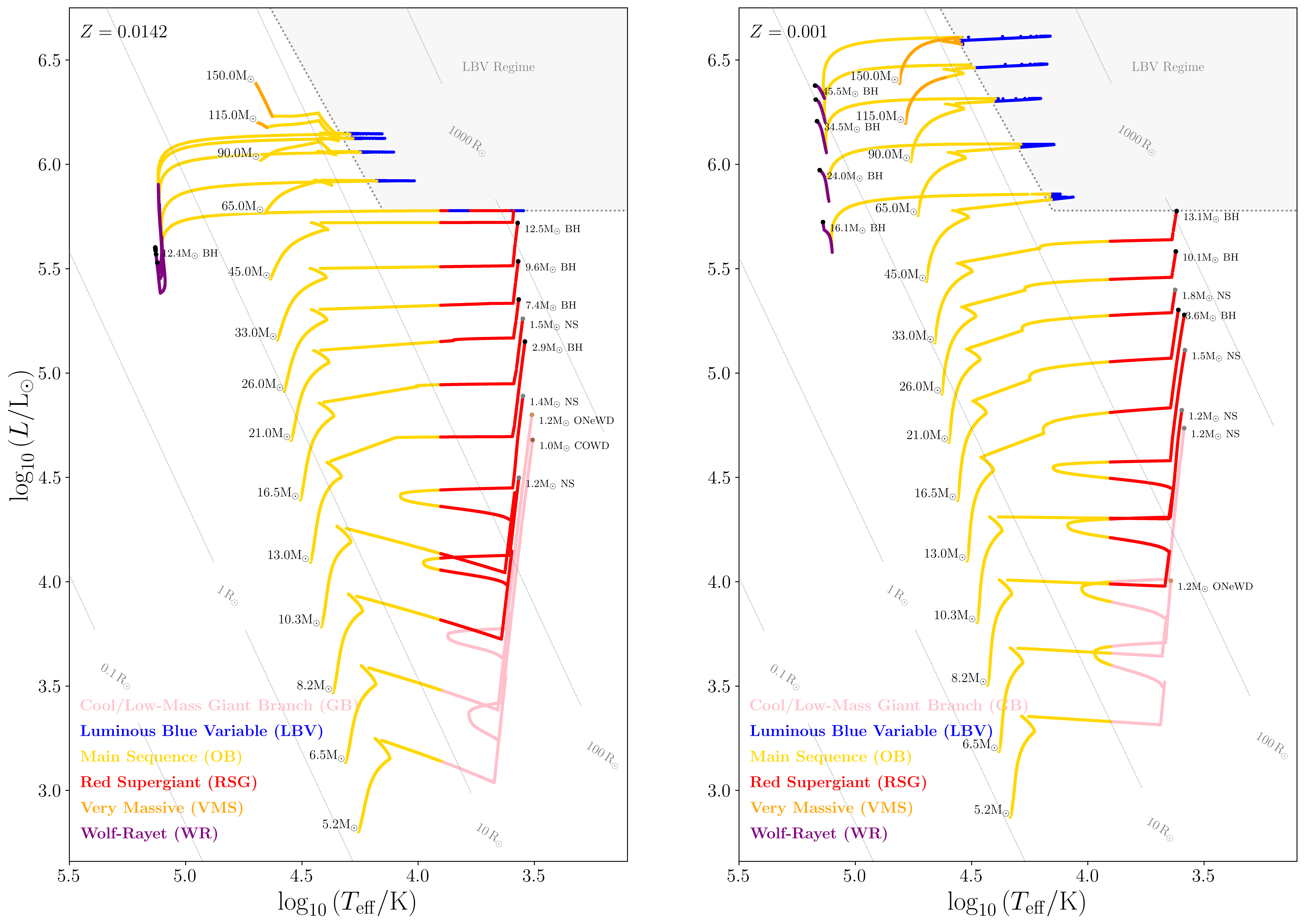}
    \caption{Hertzsprung--Russell diagram, with evolutionary tracks color-coded by the dominant mass-loss type (distinct from ``stellar type'' in COMPAS). These tracks are generated using the new \textsc{Merritt2025} combination of mass loss prescriptions.
    The left panel shows solar metallicity ($Z = 0.0142$)  tracks and the right panel shows tracks at the lower metallicity $Z = 0.001$.
    ZAMS (zero-age main-sequence) mass is annotated at the start of the track. 
    The final compact object type and mass (assuming the \citet{Mandel:2020qwb} stochastic remnant mass prescription) are also annotated at the track's end. 
The shaded region in the upper right corner of this diagram labelled `LBV regime' denotes stars that are beyond the \citet[][]{Humphreys:1979ApJ} limit and stars whose tracks enter this region are assumed to experience LBV-like mass loss (see Section~\ref{sec:methods} for details; \citealp{Hurley:2000MNRASSSE}).  
   }
    
    \label{fig:HRDDMLR}
\end{figure*}

We use the rapid binary population synthesis code COMPAS \citep[][]{Stevenson:2017tfq,COMPASTeam:2021tbl,Compas:2022JOSS} to model the evolution of massive stellar binaries.
Stellar tracks for single stars in COMPAS use the polynomial fitting formulae from \citet{Hurley:2000MNRASSSE} to the tracks from \citet{Pols:1998MNRAS}.
Binary stellar evolution is modeled following \citet{Hurley:2002MNRASBSE}, with modifications as described in \citet{COMPASTeam:2021tbl}.

In this paper, we focus on mass-loss rates for massive stars, with $M_\mathrm{ZAMS}$ (zero-age main sequence) $> 10$\,M$_\odot$, as these are the progenitors of neutron stars and black holes.
These stars have significant mass loss across the Hertzsprung-Russell diagram (HRD).

The previous wind prescription used in COMPAS is detailed in \citet{Belczynski:2010ApJ} and \citet{COMPASTeam:2021tbl}. 
In short, for massive OB main-sequence stars, mass-loss rates from \citet{Vink:2000aap,Vink:2001aap} were used. 
For cool stars ($T_\mathrm{eff} < 12,500\,\mathrm{K}$), mass-loss rates from \citet{Nieuwenhuijzen:1990aap} were used for $L>4000\,L_\odot$. For lower-mass, cooler evolved stars, winds from \citet{1993ApJ...413..641V} or \citet{KR1973} were used.
For helium-rich Wolf--Rayet stars, mass-loss rates from \citet{Hamann:1998A&A} were used, with a metallicity scaling from \citet{Vink:2005aap}, following \citet{Belczynski:2010ApJ}.
Stars that exceed the Humphreys-Davidson (HD) limit \citep[][]{Humphreys:1979ApJ} are assumed to become luminous blue variables (LBVs) and experience eruptive mass loss. 
LBV mass loss is poorly understood, short lived, and episodic. Its modeling is typically simplified in rapid binary population synthesis codes. LBV mass loss is therefore modeled as a simple time-averaged strong wind following \citet{Hurley:2000MNRASSSE}, or optionally \citet{Belczynski:2010ApJ}. {These prescriptions are designed} to ensure that the HD limit is reproduced {rather than to faithfully model LBV stars.    In our default model, we add LBV mass loss to any other relevant mass loss at the rate
\begin{eqnarray}
\dot{M}_\mathrm{LBV} &=& 0.1 \left(\frac{R}{10^{5} R_\odot} \left(\frac{L}{L_\odot}\right)^{1/2} - 1\right)^3 \nonumber \\ 
	&\times& \left( \frac{L}{6\times10^5 L_\odot} -1\right) \frac{M_\odot}{\mathrm{yr}}
\end{eqnarray}
when the HD limit is exceeded, i.e., $L > 6\times 10^5 L_\odot$ and $\frac{R}{R_\odot} \left(\frac{L}{L_\odot}\right)^{1/2} > 10^5$, following section 7.1 of \citet{Hurley:2000MNRASSSE}.}
For low-mass stars, the prescription from \citet{Hurley:2000MNRASSSE} was used.
We neglect the impact of rotation on mass loss. 
The simplified stellar models used in SSE/BSE based codes, including COMPAS, do not self consistently track the abundances of different elements over time. For mass-loss rate prescriptions that scale with metallicity, we choose to scale with the initial stellar metallicity, as is standard practice.

In the following sections, we describe the updates we have made to modeling wind mass-loss for {hot stars such as O and B main-sequence stars} (Section~\ref{subsec:OB_winds}), for very massive stars (VMS) (Section~\ref{subsec:VMS_mass_loss}), for {cool massive stars, including RSGs} (Section~\ref{subsec:RSG_mass_loss}), and for helium-rich {stripped stars and} Wolf--Rayet (WR) stars (Section~\ref{subsec:WR_mass_loss}). Additionally, we compare three sets (New Defaults/$\textsc{Merritt2025}$, $\textsc{Pessimistic}$, Old Defaults/$\textsc{Belczynski2010}$) across the four on-phase mass-loss options (OB, RSG, VMS, WR) and explore the implications on compact binary formation and rates. Choices of prescription in these sets are detailed in {tables \ref{table:prescriptions} and \ref{table:newtable}}, and the regime of each wind phase is plotted in an HR diagram in Figure \ref{fig:HRDDMLR}.

\subsection{{Hot stars}}
\label{subsec:OB_winds}

For stars with $T_\mathrm{eff} >8000\,\mathrm{K}$ and $M <  100$\,M$_\odot$, we use a set of prescriptions {that we denote as ``OB mass-loss'' as a short-hand in the text and figures, although in some cases} these are applied outside of the main sequence and outside of the spectral types O, B, including to stars early on the Hertzsprung gap (HG).  In fact, we apply these winds to all stars that satisfy the thresholds above unless they qualify as very massive stars (see Section \ref{subsec:VMS_mass_loss}), as red supergiants (see Section \ref{subsec:RSG_mass_loss}), or as naked helium stars (see Section \ref{subsec:WR_mass_loss}). Winds are explicitly set to zero for remnants (WD, NS, BH). 

Massive, main-sequence O and early B type stars experience strong line-driven winds \citep[e.g.,][]{Castor:1975ApJ,Pauldrach:1986A&A}. Based on a series of Monte Carlo simulations, \citet{Vink:2000aap,Vink:2001aap} determined a recipe for the mass-loss rates of massive OB stars as a function of stellar parameters~\footnote{We do not include an explicit scaling of the terminal wind velocity, $v_\infty$, with metallicity when using the \citet{Vink:2001aap} prescription.}
including mass, effective temperature, luminosity and metallicity.
The mass-loss rates increase for more luminous stars, meaning both that mass loss is expected to increase along the main sequence for a given star and that initially more massive stars experience stronger winds. 
Due to its driving mechanism, mass loss is further predicted to increase with metallicity. 
A discontinuity in the mass-loss behavior predicted by \citet{Vink:2000aap,Vink:2001aap} occurs at a point in effective surface temperature referred to as the ``bi-stability jump'', below which the mass loss is predicted to increase due to a recombination of iron \citep{Vink:1999aap}. 
While the precise behavior in this region is complex and still a matter of active research \citep[e.g.,][]{Petrov:2016MNRAS,Bernini-Peron:2023A&A,Krticka:2024aap}, we assume -- like most evolution codes -- a simple transition between the two mass-loss regimes at $T_\text{eff} = 25,000\,$K. 
\citet{Vink:2001aap} further predict bi-stability jumps around $15,000\,$K, and $35,000\,$K, the latter only at very low metallicity due to carbon, but we do not implement these discontinuities explicitly. 
The \citet{Vink:2021MNRAS} update to OB mass-loss provides two bi-stability jumps whose temperatures are gamma-dependent\footnote{See however \citet{deBurgos:2024A&AL} who do not find observational evidence of such an increase in mass-loss rates across the bi-stability region in a sample of Galactic massive stars. Similar results are also found in the SMC \citep[][]{Bernini-Peron:2024A&A} and LMC \citep[][]{Verhamme:2024A&A}.}.

\begin{figure}
    \centering

    \includegraphics[width=1.0\columnwidth]{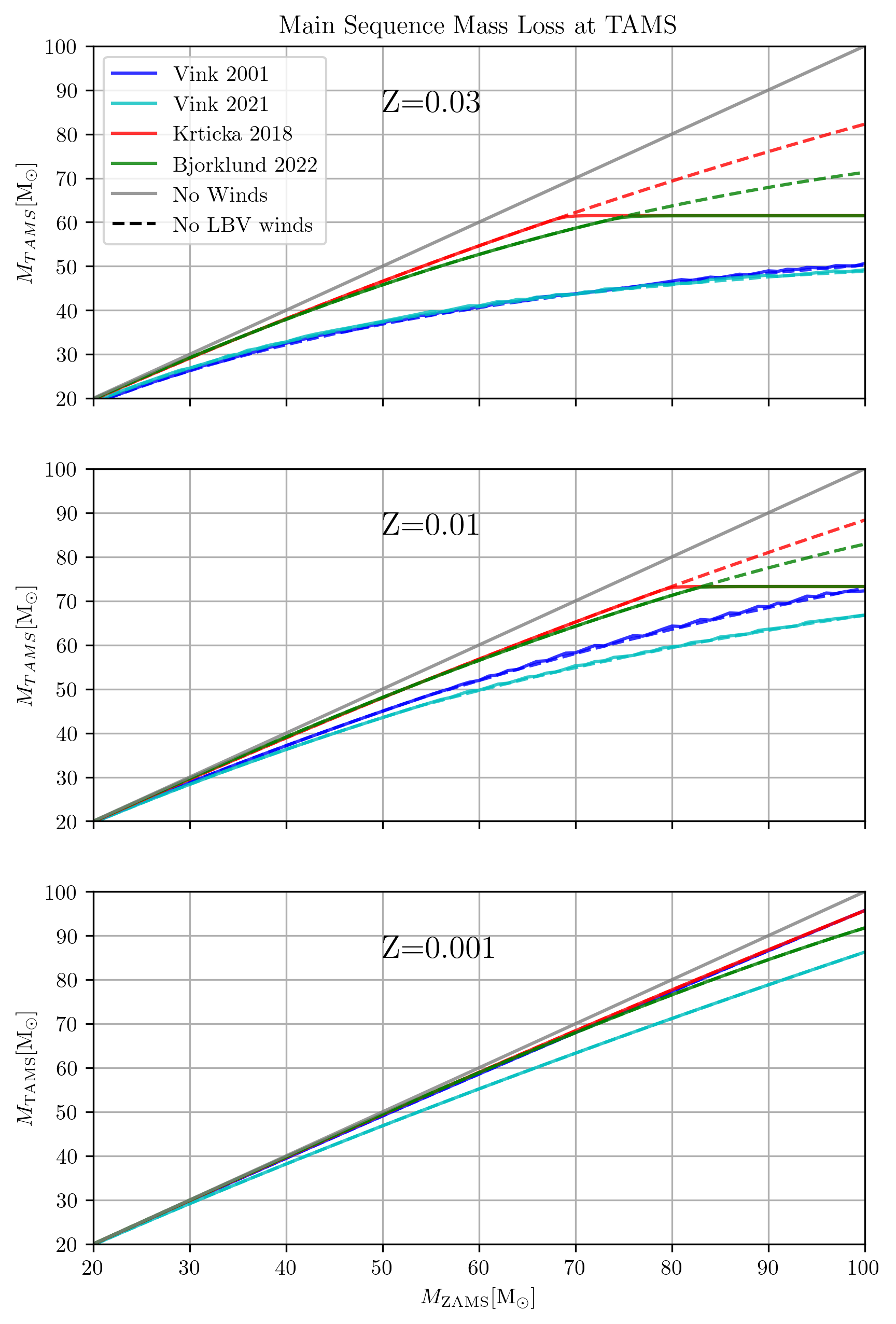}

    \caption{Terminal-age main-sequence (TAMS) mass $M_\mathrm{TAMS}$ as a function of the initial zero-age main-sequence (ZAMS) mass $M_\mathrm{ZAMS}$.
    The three panels show the results for different metallicities: Z=[0.03, 0.01, 0.001].
    The blue curve uses the \citet{Vink:2001aap} prescription, the cyan curve uses the \citet{Vink:2021MNRAS} prescription, the red curve uses the \citet{Krticka:2018aap} prescription and the green curve uses the \citet{Bjorklund:2023aap}.
    The previous default prescription is \citet{Vink:2001aap}. 
    All prescriptions exhibit a strong metallicity dependence.
    The gray diagonal line indicates no main-sequence mass loss. 
    Stars that exceed the Humphreys--Davidson limit may experience additional LBV mass loss already during core H burning, leading to a plateau at high mass; dashed lines have LBV winds turned off to demonstrate only the impact of main-sequence mass loss.
    }
    \label{fig:MZAMS_MTAMS}
\end{figure}

While the direct mass removal on the main sequence is usually low compared to the subsequent evolution stages, the amount of main-sequence mass loss can considerably influence which subsequent regimes are reached and how much time stars will spend in them \citep[e.g.,][]{Langer:1995A&A,Higgins:2020MNRAS,Josiek:2024arXiv}.
In COMPAS, stars that lose mass during the main sequence transition to a stellar track of a non-mass-losing star of the reduced mass, at a time that preserves the fractional main-sequence lifetime.  This simplified model, which does not track the composition and structure of the partially stripped star, can lead to errors in the stellar luminosity, which may affect subsequent mass evolution predictions \citep[][]{shikauchi2024evolutionconvectivecoremass}.

Recently, several groups have determined the mass-loss rates for massive OB stars, finding that they are typically lower than those found by \citet{Vink:2000aap,Vink:2001aap}.
Using non-thermodynamic equilibrium radiative transfer in the co-moving frame to connect the photosphere to the wind, \citet{Krticka:2017aap, Krticka:2018aap} calculate global, unified models of both the stellar atmosphere and the wind for O type dwarf, giant and supergiant stars at a range of metallicities\footnote{We do not incorporate the recent extension of this model to a wider parameter space including lower metallicities \citep[][]{Krticka:2021A&A,Krticka:2024aap,Krticka:2025A&A}; we leave incorporating these more recent updates to future work.}. 
Their models predict lower mass-loss rates than \citet{Vink:2000aap,Vink:2001aap}. 
This is in line with the general trend towards lower mass-loss rates due to the presence of wind clumping \citep[e.g.,][]{Fullerton:2006ApJ,Muijres:2011A&A,Vink:2022ARA&A}, although the empirical picture remains complex \citep[e.g.,][]{Ramachandran:2019A&A,Brands:2022A&A,Rickard:2022A&A} with post-interaction binaries and other evolved objects yielding higher mass-loss rates \citep[e.g.,][]{Bouret:2021A&A,Pauli:2022A&A,Pauli:2023A&A} and thus blurring any sample of main-sequence OB stars.
\citet{Vink:2021MNRAS} update the Monte Carlo wind simulations of \citet{Vink:2000aap,Vink:2001aap} and include a consistent calculation of the wind terminal velocity \citep[following][]{Mueller:2008A&A}, finding a shallower dependence of the mass-loss rate on metallicity than their previous models. 
Similar to \citet{Krticka:2017aap,Krticka:2018aap}, \citet{Bjorklund:2023aap} solve the spherically symmetric steady-state equation of motion using non-thermodynamic equilibrium radiative transfer in the co-moving frame for a grid of models. 
They provide a fit for the mass-loss rates of massive O and B-type stars as a function of stellar parameters, finding that their rates are significantly lower than those of \citet{Vink:2000aap,Vink:2001aap}.
In addition, \citet{Bjorklund:2023aap} do not find a significant increase in mass-loss rates in the bi-stability region, contrary to \citet{Vink:2000aap,Vink:2001aap}. 
This may be in line with recent observational evidence against the existence of the bistability jump \citep[][]{deBurgos:2024A&AL,Bernini-Peron:2024A&A,Verhamme:2024A&A}.
We implement the updated mass-loss prescriptions derived by \citet{Krticka:2018aap}, \citet{Vink:2021MNRAS} (new default) and \citet{Bjorklund:2023aap} in COMPAS.

For massive stars with initial, zero-age main-sequence (ZAMS) masses $10 < M_\mathrm{ZAMS} / M_\odot < 100$ at three different metallicities ($Z = [0.001, 0.01, 0.03]$), we show the terminal-age main-sequence (TAMS) mass $M_\mathrm{TAMS}$ predicted by each of these prescriptions in Figure~\ref{fig:MZAMS_MTAMS}.   The \citet{Krticka:2018aap} and \citet{Bjorklund:2023aap} prescriptions produce similar results, both giving less than half the main-sequence mass loss compared to \citet{Vink:2001aap}.
We also find that at high metallicity ($Z \gtrsim 0.01$) the implementation of both the \citet{Krticka:2018aap} and \citet{Bjorklund:2023aap} prescriptions in COMPAS results in a plateau in $M_\mathrm{TAMS}$ as a function of $M_\mathrm{ZAMS}$ (cf.\ Fig.\,\ref{fig:MZAMS_MTAMS}). This plateau arises as stars that exceed the
Humphreys–Davidson (HD) limit \citep[][]{Humphreys:1979ApJ} may experience additional LBV
mass loss already during core H burning, leading to a plateau
at high mass; dashed lines have LBV mass-loss turned off to
demonstrate the impact of the implemented main-sequence
wind prescriptions.

\subsection{Very massive stars}
\label{subsec:VMS_mass_loss}

Very massive stars \citep[here with $M \geq 100$\,M$_\odot$, following][]{Vink:2015HiA} approach the Eddington limit while still core hydrogen burning.  In COMPAS, we compute the Eddington parameter $\Gamma_\text{e}$ as 

\begin{equation}
    \Gamma_\text{e} = 2.49 \times 10^{-5} \left(\frac{L}{\text{L}_\odot}\right)\left(\frac{M}{\text{M}_\odot}\right)^{-1}.
    \label{eq:EddingtonFactorImplemented}
\end{equation}

The mass loss rate increases rapidly as the Eddington parameter approaches unity \citep{Vink:2011aap,Bestenlehner:2014aap}. 
The origin of this behavior is a transition from optically thin to optically thick winds \citep{Vink:2012ApJ,Sabhahit:2023lni}, and corresponds to a spectral type transition from O to Of(Evolved O-type with nitrogen and helium emission lines)/WNh(Wolf-Rayet of the nitrogen sequence with hydrogen), and eventually to WNh \citep{Crowther:2011MNRAS}. 

\citet{Vink:2011aap} show the difference between their predictions and those of \citet{Vink:2000aap,Vink:2001aap}, as a function of $\Gamma_\text{e}$ in their Figure~5.
We model the prediction of \citet{Vink:2011aap} by adding a quadratic polynomial in $\Gamma_\text{e}$ to the logarithm\footnote{All logs are base 10.} of the \citet{Vink:2000aap,Vink:2001aap} mass-loss rate for stars with $\Gamma_\text{e} > 0.5$:
\begin{equation}
    \log \dot{M}_\mathrm{V11} = \log \dot{M}_\mathrm{V01} + F(\Gamma_\text{e}) ,
    \label{eq:logMdotV11}
\end{equation}
 where $\log \dot{M}_\mathrm{V01}$ is the mass-loss rate according to \citet{Vink:2000aap,Vink:2001aap} and 
\begin{equation}
    F(\Gamma_\text{e}) = a_0 + a_1 \Gamma_\text{e} + a_2 \Gamma_\text{e}^2 
    \label{eq:FGammaV11Quadratic}
\end{equation}
with $a_0 = 0.0447$, $a_1 = 0.309$ and $a_2 = 0.243$. The accuracy of this fit can be quantified with an $r^2$ value of 0.983.

\citet{Sabhahit:2023lni} recently presented a further improvement of this model involving a $\Gamma_\text{e}$-dependent switch point based on a set of MESA models, below which the low efficiency of winds dictates the use of the \citet{Vink:2001aap} prescriptions, and above which the VMS \citet{Vink:2011aap} prescription is used. 
The switch point occurs at a metallicity-dependent mass, luminosity, and escape velocity. 
They suggest implementing an iterative search algorithm to determine the switch point at each time step.
However, for the speed desired in many population synthesis applications, we implement the prescription using simple fits to their switch mass and luminosity by fitting the data in their Table 2 as
\begin{equation}
    \Gamma_\mathrm{SWITCH} = 2.49 \times 10^{-5} \cdot 
    \frac{L_\mathrm{SWITCH}}{\text{L}_\odot} 
    \cdot
    \frac{\text{M}_\odot}{M_\mathrm{SWITCH}} ,
    \label{eq:Sabhahit2023_GammaSwitch}
\end{equation}
where
\begin{equation}
    L_\mathrm{SWITCH} / \text{L}_\odot = 10^{2.36} Z^{-1.91},
    \label{eq:Sabhahit2023_LSwitch}
\end{equation}
and
\begin{equation}
    M_\mathrm{SWITCH} / \text{M}_\odot = 0.0615 Z^{-1.574} + 18.10 . 
    \label{eq:Sabhahit2023_MSwitch}
\end{equation}
Finally, if $\Gamma_\text{e} > \Gamma_\mathrm{SWITCH}$, we model the mass-loss rate as
\begin{equation}
    \dot{M}_\mathrm{Sabhahit2023} = \dot{M}_\mathrm{SWITCH} \left( \frac{L}{L_\mathrm{SWITCH}} \right)^{4.77} 
    \left( \frac{M}{M_\mathrm{SWITCH}} \right)^{-3.99} ,
    \label{eq:Sabhahit2023_Mdot}
\end{equation}
where
\begin{equation}
    \log (\dot{M}_\mathrm{SWITCH} / \text{M}_\odot\,\rm{yr}^{-1}) = -1.86 \log(Z) - 8.90;
    \label{eq:Sabhahit2023_MdotSwitch}
\end{equation}
otherwise, if $\Gamma_e \leq \Gamma_\mathrm{SWITCH}$, $\dot{M}_\mathrm{Sabhahit2023}$ follows the prescription for OB mass loss, for which  \citet{Vink:2021MNRAS} is the default model.

Finally, we also include the VMS mass-loss prescription from \citet{Bestenlehner:2020MNRAS}. 
This analytic prescription is a modified form of the original line driven theory from \citet{Castor:1975ApJ} based on a study of the transition in $\Gamma_\text{e}$ from optically thin to thick winds near $\Gamma_e \sim 0.5$.

\begin{figure*}
    \centering
    \includegraphics[width=\textwidth]{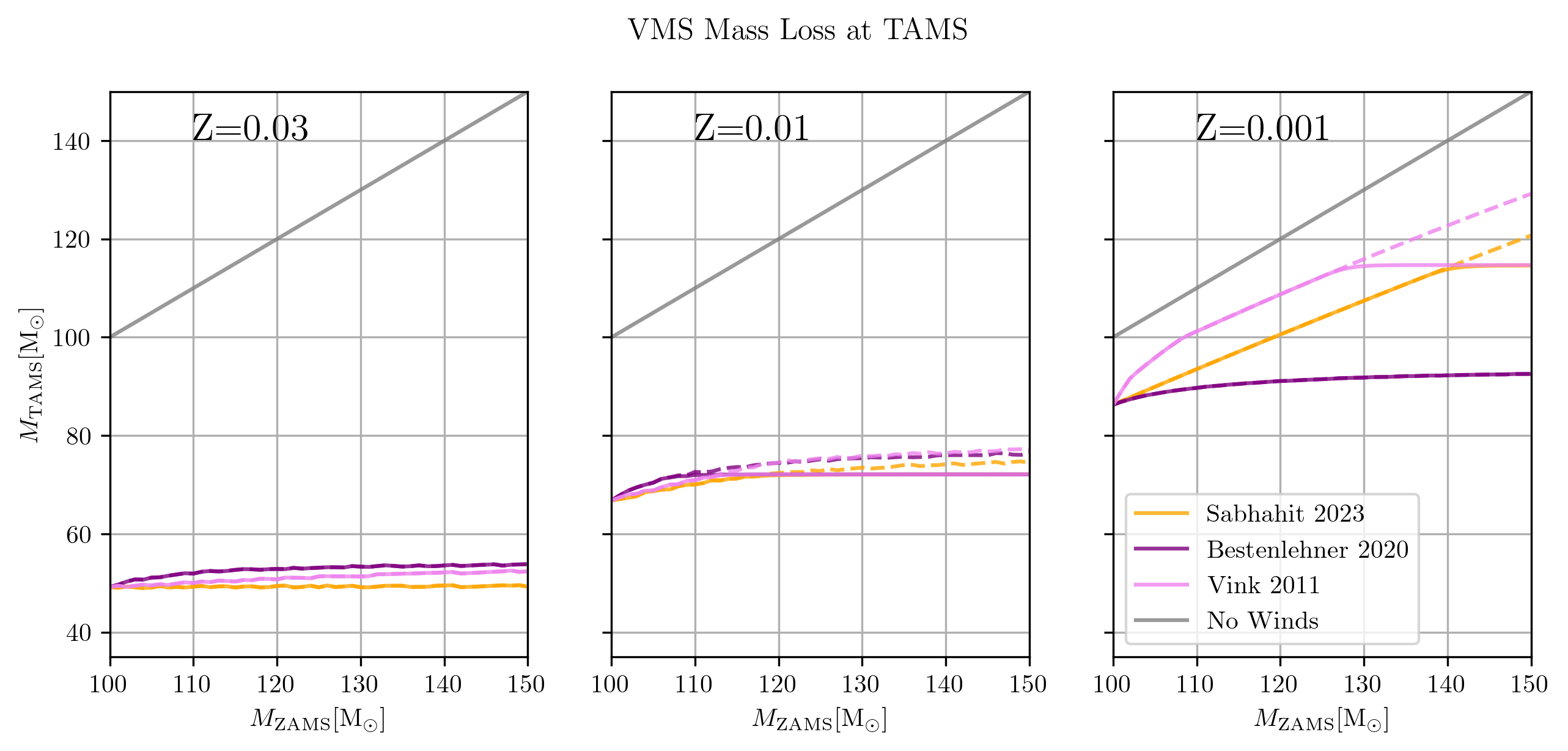}
    \caption{$M_\mathrm{TAMS}$ as a function of $M_\mathrm{ZAMS}$  for the three VMS mass-loss prescriptions described in Section~\ref{subsec:VMS_mass_loss}. The three panels show the results for different metallicities: $Z=[0.03, 0.01, 0.001]$.  
    The dotted lines show how VMS mass-loss treatment would proceed without the onset of LBV winds above the HD limit.
    The gray diagonal line shows $M_\mathrm{TAMS} = M_\mathrm{ZAMS}$, for no mass loss on the MS. 
    }
    \label{fig:MTAMS-MZAMS-VMS} 
\end{figure*}

In Figure~\ref{fig:MTAMS-MZAMS-VMS} we show $M_\mathrm{TAMS}$ as a function of $M_\mathrm{ZAMS}$ in the regime where VMS winds occur for each prescription at three choices of metallicity ($Z=[0.03, 0.01, 0.001]$). At high (0.03) and near solar (0.01) metallicity, VMS with a broad range of initial masses yield similar $M_\mathrm{TAMS} \sim 50$\,M$_\odot$, and $M_\mathrm{TAMS} \sim 75$\,M$_\odot$, respectively, for all three VMS mass loss prescriptions, as all stars with $M_\mathrm{ZAMS} \leq 150$\,M$_\odot$ drop below the mass threshold for VMS winds for all three VMS prescriptions while still on the main sequence. The choices of VMS prescription diverge significantly in TAMS mass at low metallicity (0.001), however. \citet{Sabhahit:2023lni} exhibits the greatest metallicity dependence, while \citet{Bestenlehner:2020MNRAS} exhibits the least.

\subsection{{Cool massive stars}}
\label{subsec:RSG_mass_loss}

Below $T_\mathrm{eff} < 8000\,\mathrm{K}$, line-driven winds are no longer considered to be efficient and the mass loss mechanism changes. 
Stars with $8000\,\mathrm{K} > T_\text{eff} > 3500\,$K are considered ``yellow'' supergiants or hypergiants \citep{deJager:1998A&ARv}, but the mass loss in this sparsely populated regime---also termed the ``yellow void'' \citep[e.g.,][]{Nieuwenhuijzen:2000A&A}---is highly uncertain with pulsations and outbursts possibly being more important than continuous winds for these stars \citep[e.g.,][]{Koumpia:2020A&A,Humphreys:2023AJ}. Below $T_\mathrm{eff} \sim 3500\,\mathrm{K}$, we reach the regime of red supergiants, which is much more populated and much better studied. 
The environment of the stars in this regime is usually cool enough to allow dust formation, which is thought to play a major role in explaining the winds of red supergiants -- hence their common terminology as ``dust driven'' -- though still with considerable uncertainties about the actual mechanisms and the strength of the outflows \citep[see, e.g.,][for a recent review]{Decin:2021ARA&A}. 
Given the absence of dedicated recipes for the yellow supergiant domain, we follow the common strategy of many stellar evolution codes and treat the regime of evolved yellow and red massive stars with the same description, which for simplicity we call ``red supergiants'' (RSGs) for the rest of this work.  We thus apply {these winds to all cool massive stars, i.e.,} giant stars with a hydrogen-rich envelope with $T_\text{eff} < 8000\,$K and ZAMS mass above 8 $\text{M}_\odot$ {(we use the short-hand ``RSG mass-loss'' in the text and figures)}.  Less massive cool stars, labeled ``Cool/Low-Mass Giant Branch'' in Fig.~\ref{fig:HRDDMLR}, experience \citet{Nieuwenhuijzen:1990aap}  winds.

RSGs are observed as the progenitors of some hydrogen-rich type-IIP supernovae \citep[][]{Smartt_2009,Smartt:2015PASA}. 
The maximum observed RSG luminosity is around {$\log (L/\text{L}_\odot ) \approx 5.5$} \citep[][]{Davies_2020}, and the maximum luminosity of a type-IIP RSG progenitor is fainter at $\log L/\text{L}_\odot \approx 5.1$, leading to the so called missing red supergiant problem \citep[][]{Smartt_2009}.\footnote{However, the significance of the missing RSGs is now estimated to be less than $2\sigma$ \citep[][]{Davies_2020}, so there may not, in fact, be a missing population of luminous RSG type-IIP progenitors.}  There are several proposed explanations for this apparent discrepancy.

One proposal is that above some mass threshold (now estimated at $19^{+4}_{-2}\,\text{M}_\odot$), the missing RSGs experience a type of yet-unobserved failed supernova\footnote{Though see \citet{Neustadt:2021MNRAS} for potential candidates.} and directly collapse to form a black hole. This proposal has gained support from numerical simulations \citep{O_Connor_2011}, and motivated searches for the disappearance of RSGs, with possible but inconclusive candidates found (\citealt{Adams_2017}, but see also \citealt{Beasor:2024}).

Another possible explanation of the missing RSG problem is that massive RSGs lose their hydrogen envelopes through mass loss, end their lives as stripped Wolf--Rayet stars, and produce hydrogen-poor (type I) supernovae \citep[e.g.,][]{Yoon:2010ApJL,Georgy:2012A&A}. 
If a RSG undergoes core collapse and explodes with a hydrogen envelope, it will appear as a type-IIP supernova. 
If it loses its envelope through mass loss and becomes a Wolf--Rayet/helium star, it would go on to produce a H-poor supernova if it explodes. 
We find that these outcomes are very sensitive to the chosen mass-loss prescription during the RSG phase, and have major implications for the predicted ratios of different SN types and for the mass distributions of compact remnants. 

Typically, stellar evolution and population synthesis models have employed empirical prescriptions for RSG mass loss. 
The previous default in COMPAS \citep{deJager:1988aapS,Nieuwenhuijzen:1990aap} is a function of metallicity, luminosity, mass, and radius based on a relatively small sample (271 stars) across the entire HR diagram, \textit{of which only 14 were RSGs}.  
Recently, a number of updated RSG mass-loss prescriptions have become available.
There was no ``RSG'' specific designation in mass-loss or stellar type, rather \citet{deJager:1988aapS,Nieuwenhuijzen:1990aap} was applied to stars with $T < 8000$\,K and $L > 4000\,\text{L}_\odot$. Our new switching criteria provide a set of RSG specific mass loss options, and applies to stars with $T < 8000$\,K (thus including YSG and cooler RSG), $M_\mathrm{ZAMS} >8$\,M$_\odot$, and belonging to one of the following evolved stellar types: Hertzsprung Gap (HG), Core Helium Burning (CHeB), First Giant Branch (FGB), Early Asymptotic Giant Branch (EAGB), Thermally Pulsing Asymptotic Giant Branch (TPAGB).

\citet{Beasor:2020MNRAS,Beasor:2023MNRAS} derived an empirical RSG mass-loss prescription based on observations of RSGs in clusters, allowing them to fit the mass loss as a function of stellar mass $M$ and luminosity $L$. 
This results in significantly lower mass-loss rates for RSGs compared to \citet{Nieuwenhuijzen:1990aap}, with even the most massive RSGs considered (25\,M$_\odot$) by \citet{Beasor:2020MNRAS,Beasor:2023MNRAS} retaining their hydrogen-rich envelopes (losing less than 1\,M$_\odot$ during the RSG phase) and avoiding self-stripping completely.

\citet{Kee:2021aap} provide an analytical description of RSG mass loss by focusing on turbulence as the dominant mass-loss mechanism, and then further correct it numerically. 
This prescription is included in COMPAS for its unique approach, however, it proves very sensitive to the turbulent velocity parameter. 
While this parameter is somewhat constrained from observation and from theory, within a reasonable range (15--21\,km/s) the mass-loss rates vary by four orders of magnitude, limiting the predictive power of this prescription.

\begin{figure}
    \centering
    \includegraphics[width=\columnwidth]{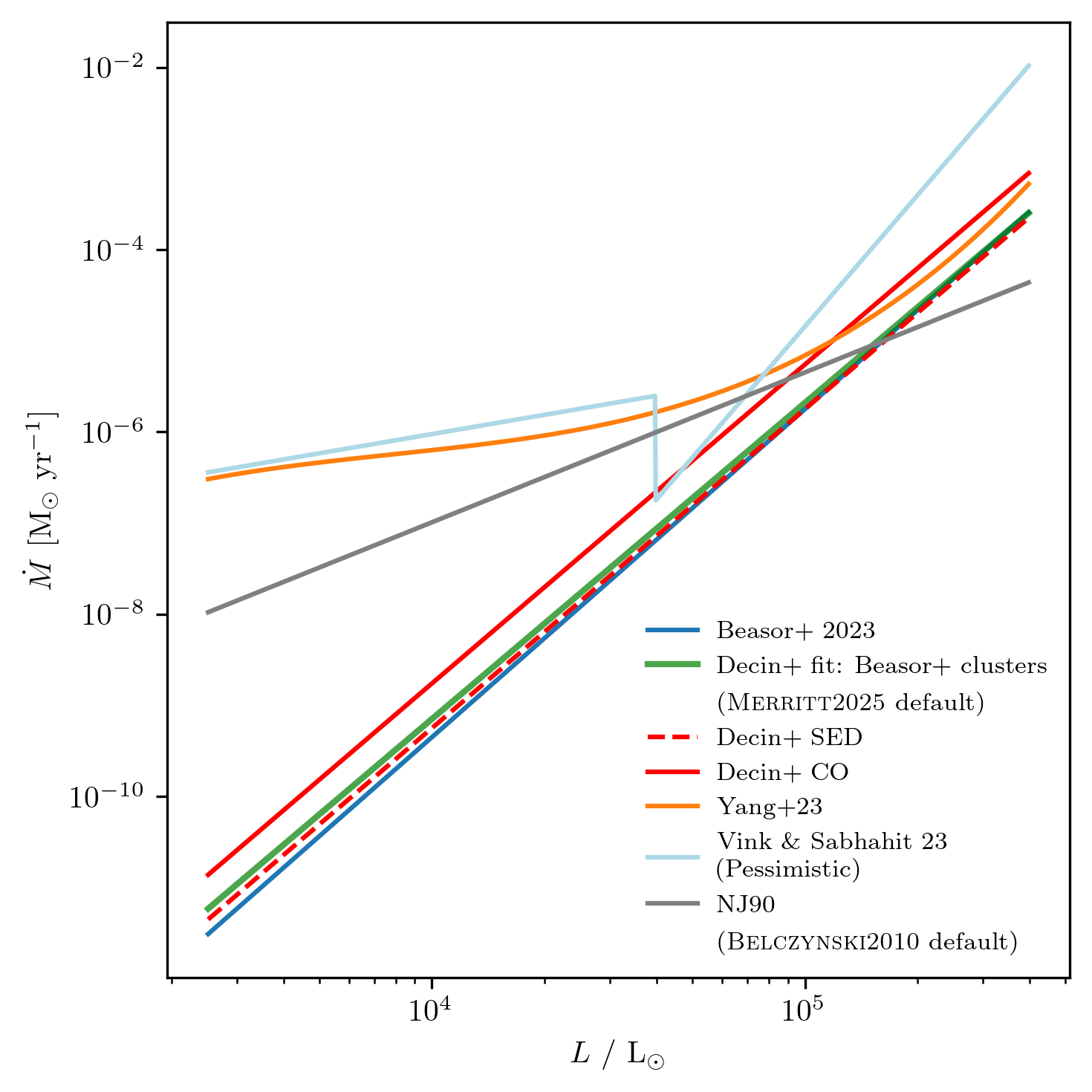}
    \caption{ 
Mass-loss rate as a function of luminosity for {cool massive stars (RSGs)} according to the various prescriptions discussed in this paper. 
    For prescriptions that are dependent on mass, we assume a mass of $15$\,M$_\odot$ for visualization purposes. For \citet{Nieuwenhuijzen:1990aap} winds (labeled NJ90), solar metallicity and a surface temperature of 4000\,K is assumed. \citet{Kee:2021aap} is not plotted, because rates vary over four orders of magnitude given reasonable choices of the turbulent velocity.}
    \label{fig:MdotRSG}
\end{figure}

\citet{Decin:2024AA} use ALMA data on CO line emission to determine the mass-loss rates of five red supergiants in the open cluster RSGC1. 
They find mass-loss rates systematically higher than those determined via SED fitting (solid (CO) vs. dashed (SED) red lines in Figure \ref{fig:MdotRSG}).
\citet{Decin:2024AA} then use their ALMA data to recalibrate the mass-loss rates derived by \citet{Beasor:2020MNRAS} for RSGs in several clusters (solid green line in Figure \ref{fig:MdotRSG}). We take this as our default RSG mass-loss prescription in COMPAS. 
\citet{Decin:2024AA} find that with this mass-loss prescription, RSGs do not lose their hydrogen envelopes due to wind mass loss. 

\citet{Yang:2023arXiv} study the mass-loss rates of a large sample of RSGs observed in the Small Magellanic Cloud (SMC). They obtain a third-order polynomial fit in $L$ for the mass-loss rate, and find rates that are markedly higher than \citet{Beasor:2020MNRAS,Beasor:2023MNRAS} or \citet{Decin:2024AA}, particularly for low luminosities ($L < 10^5\,\text{L}_\odot$).

\citet{VinkSabhahit:2023aap} reinterpret the data from \citet{Yang:2023arXiv}, identifying a kink in the mass-loss rate at a luminosity of $\log (L / \text{L}_\odot) = 4.6 $, and associating this with the onset of multiple scattering.
They fit a functional form with both $L$ and $M$ dependence, similar to \citet{Beasor:2020MNRAS,Beasor:2023MNRAS} and \citet{Decin:2024AA}.
With this prescription, \citet{VinkSabhahit:2023aap} recover the Humphreys-Davidson limit \citep[][]{Humphreys:1979ApJ} at high luminosity, finding that high-mass RSGs lose their hydrogen envelopes, providing a solution to the missing red supergiant problem \citep[][]{Smartt_2009}. 

\begin{figure}
    \centering
    \includegraphics[width=\columnwidth]{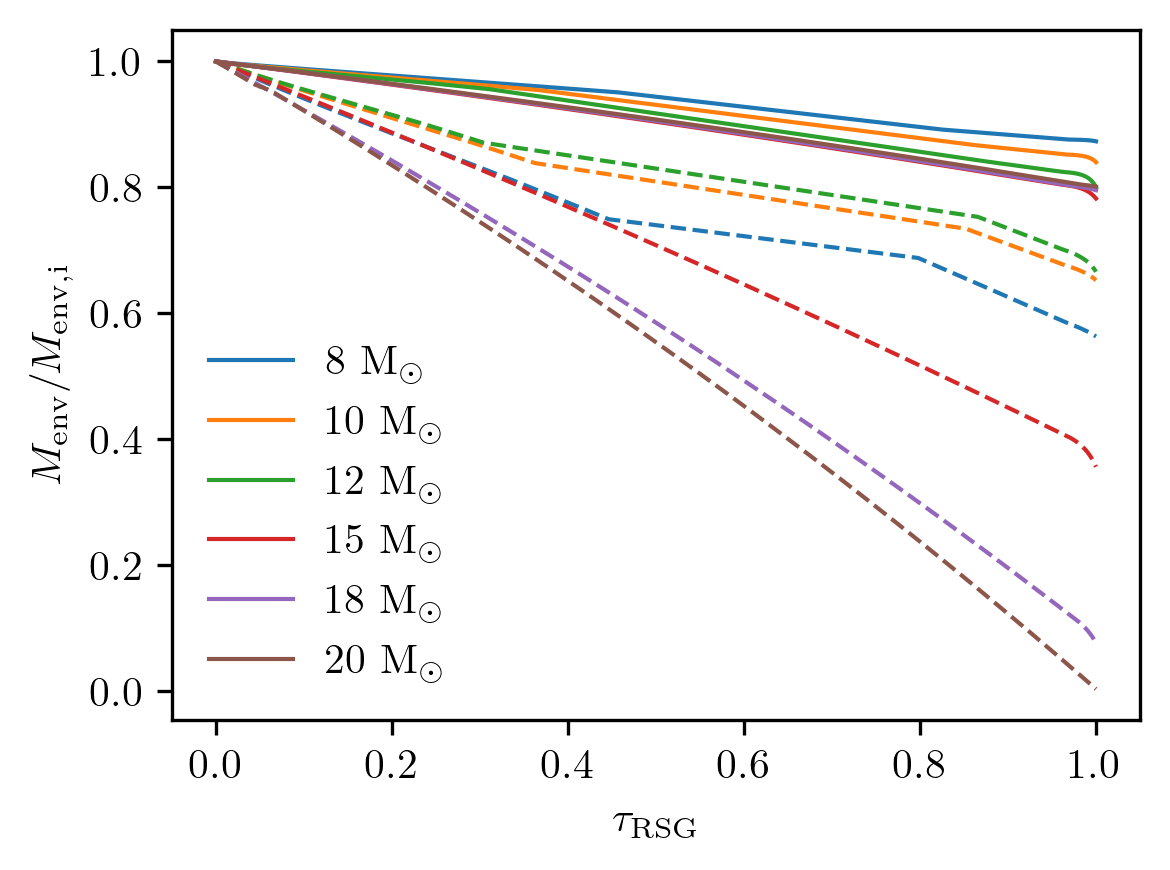}
    \caption{Fraction of envelope mass remaining for massive stars during the RSG phase. The abscissa is the fractional time along the RSG phase, while the ordinate is normalized to the envelope mass at the start of the RSG phase.
    The solid lines use our standard RSG prescription \citep[][]{Decin:2024AA}, while the dashed curves use the prescription from \citet{Yang:2023arXiv}. The colors denote the initial (ZAMS) mass of the stars in solar masses, as labeled in the legend. Solar metallicity is assumed.
    } 
    \label{fig:RSGEnvelopeMassLoss}
\end{figure}

We compare the impact of the weak RSG mass loss of \citet{Decin:2024AA} to the strong RSG mass loss of \citet{Yang:2023arXiv} in Figure~\ref{fig:RSGEnvelopeMassLoss}. We show the fraction of hydrogen envelope mass remaining as a function of the dimensionless fractional time along the RSG phase $\tau_\mathrm{RSG} = (t_\mathrm{RSG} - t_\mathrm{RSG,i} ) / (t_\mathrm{RSG,f} - t_\mathrm{RSG,i})$, where $t_\mathrm{RSG}$ is the age of the star on the RSG phase and subscripts $i$ and $f$ denote the start and end of the RSG phase.
The \citet{Decin:2024AA} prescription predicts that RSGs of masses 8--20\,M$_\odot$ retain a significant fraction of their hydrogen envelope. The much stronger mass-loss in the \citet{Yang:2023arXiv} prescription leads RSGs with masses $\gtrsim 18$\,M$_\odot$ to lose their envelopes and self strip, resulting in no massive RSG progenitors to type IIp supernovae. 

We have implemented the \citet{Beasor:2020MNRAS,Beasor:2023MNRAS}, \citet{Decin:2024AA}, \citet{Yang:2023arXiv} and \citet{VinkSabhahit:2023aap} RSG mass-loss prescriptions into COMPAS. 
Figure~\ref{fig:MdotRSG} shows a comparison of the mass-loss rates for RSGs predicted by each of these prescriptions. 
We choose the \citet{Decin:2024AA} prescription as our new default mass-loss rate. 

While the red supergiant populations are not necessarily the same at different metallicities \citep[e.g.,][]{Bonanos:2023arXiv,Ou:2023ApJ,deWit:2024arXiv}, the presumed mechanisms based on pulsation and dust as well as observational constraints suggest that the individual mass-loss rates may be largely insensitive to metallicity \citep[e.g.,][though see \citealt{Mauron:2011A&A}]{Goldman:2017MNRAS,Antoniadis:2024arXiv}. Moreover, the maximum RSG luminosity appears to be independent of metallicity \citep[at least for $Z > 0.25\,\text{Z}_\odot$;][]{McDonald:2022MNRAS} suggesting that the mechanism leading to the Humphreys-Davidson limit \citep[][]{Humphreys:1979ApJ} is not strongly metallicity dependent. By default, we do not scale RSG mass-loss rates with metallicity. 

Analyzing a large sample of RSGs in M31 and M33, \citet{Wang:2021ApJ} find there can be an order-of-magnitude difference in the mass-loss rates between RSGs with carbon-rich dust (which typically have lower mass-loss rates) and those with oxygen/silicate rich dust.
The inferred mass-loss rate from spectral energy distribution fitting also depends on the assumed gas-to-dust ratio \citep[][]{Wang:2021ApJ}. We have not implemented their prescription here, but aim to implement it in a future version of COMPAS. 

Ultimately, we find that there are orders-of-magnitude variations between the different RSG mass-loss prescriptions, such that the mass loss of massive RSGs remains highly uncertain. 

\subsection{Stripped helium star mass loss}
\label{subsec:WR_mass_loss}

Massive naked helium stars can arise either from high-mass single stars through wind stripping, from binary stars stripped through binary interactions, or a combination of both \citep[][]{Abbott:1987ARA&A,Vanbeveren:1998A&ARv,Crowther:2007ARA&A,Shenar:2020aap}.
Classical Wolf--Rayet (WR) stars are massive, evolved, helium-rich stars, which lose mass at high rates through optically thick winds. Formally defined by their spectral appearance, here we use the term WR to refer to naked helium stars, as other stars that may have WR-type spectra (e.g., VMS) are subject to a different wind treatment in COMPAS. 
In this section we describe mass-loss rates applied to stars that are fusing helium in their cores on the helium main sequence (HeMS), as well as stars on the helium Hertzsprung gap (HeHG) and helium giant branch (HeGB). 

Similar to RSGs, WR mass-loss prescriptions are typically empirical. Due to their optically thick winds, the mass of WR stars is not directly accessible from their spectra, meaning that resulting empirical recipes mainly scale with the luminosity of the stars \citep[e.g.,][]{Nugis:2000A&A,Hamann:2006A&A,Sander:2014A&A,Tramper:2016ApJ,Shenar:2019aap}.
In our previous model for WR winds in COMPAS, we followed the prescription from \citet[][]{Belczynski:2010ApJ}, which reduces the mass-loss rates from \citet{Hamann:1995A&A} by a factor of 10 \citep[motivated by][]{Yoon:2006A&A}, and applies a scaling of the mass-loss rate with initial metallicity proportional to $Z^{0.86}$, from \citet{Vink:2005aap}. 

\begin{figure}
    \centering
    \includegraphics[width=\columnwidth]{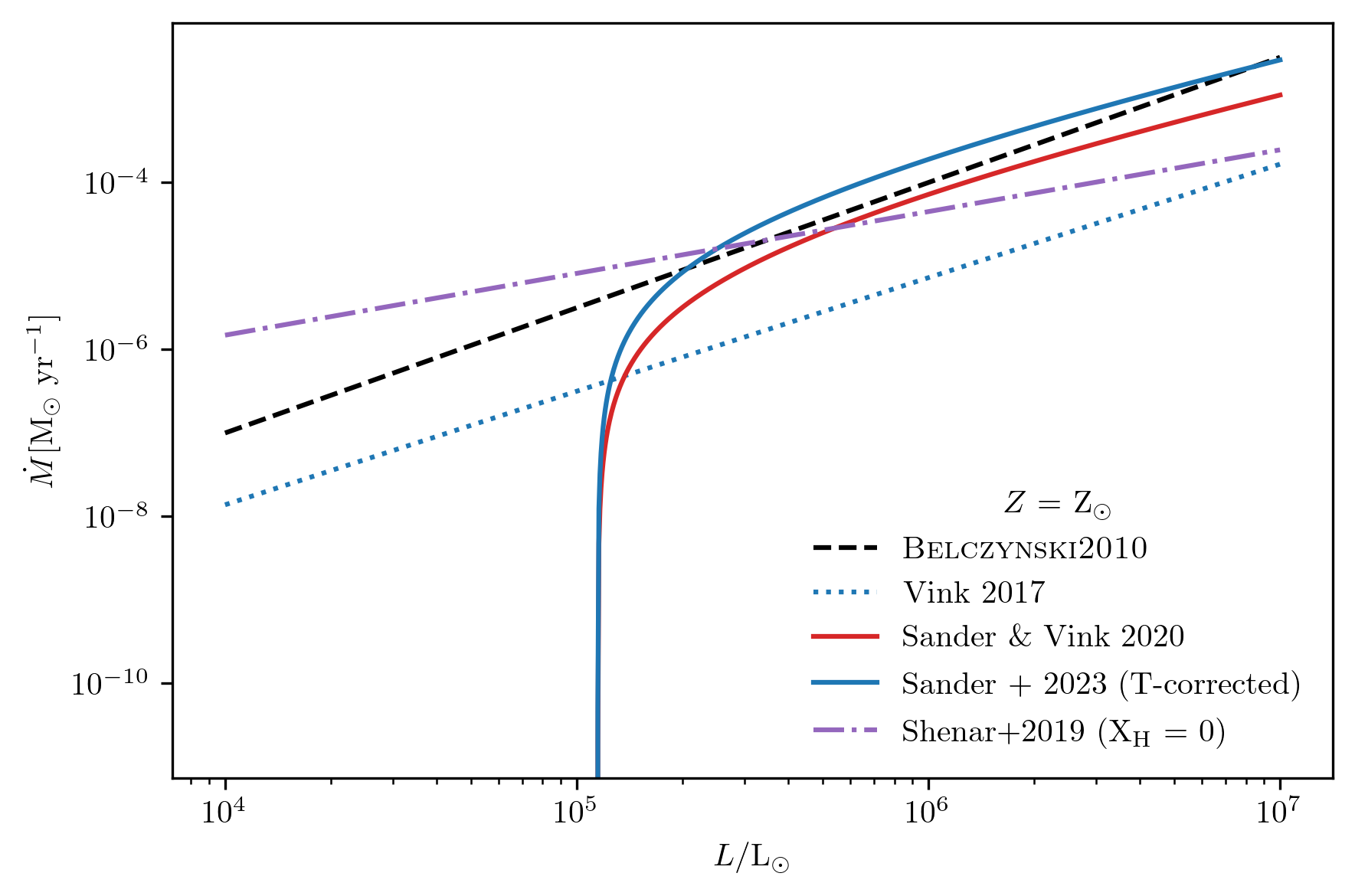}
    \caption{Mass-loss rates for helium-rich stripped stars and WR stars as a function of luminosity at solar metallicity. The dashed black line shows the previous prescription in COMPAS \citep{Hamann:1995A&A,Vink:2005aap,Belczynski:2010ApJ}.
    The dotted blue curve shows the mass-loss prescription from \citet{Vink:2017aap}. 
    The solid red curve shows the mass-loss prescription for massive WR stars from \citet{Sander:2020MNRAS}, without applying the temperature correction from \citet{Sander:2023aap}. 
    As discussed in Section~\ref{subsec:WR_mass_loss}, in COMPAS we use the maximum of the \citet{Vink:2017aap} prescription and the \citet{Sander:2020MNRAS} prescription with the temperature correction applied (cf. Equation~\ref{eq:mdot_WR}). The blue curve shows the correction applied at an effective temperature of 120,000 K.
    The dashed purple curve shows the \citet{Shenar:2019aap} mass-loss prescription.
}
    \label{fig:WR_mass_loss_rates}
\end{figure}

Recently, \citet{Sander:2020MNRAS} presented the first thorough calculations of WR mass-loss rates using dynamically-consistent stellar atmosphere models \citep{Sander:2017aap}. 
The atmosphere models of \citet{Sander:2020MNRAS} made use of a single fixed effective temperature. Subsequently, \citet{Sander:2023aap} derived a temperature correction to the mass-loss rates presented in \citet{Sander:2020MNRAS} accounting for a realistic range of WR effective temperatures. We account for the temperature correction as
\begin{equation}
    \log \left( \frac{\dot{M}_\mathrm{SV2023}}{\mathrm{M_\odot\,yr}^{-1}} \right) = \log \left( \frac{\dot{M}_\mathrm{SV2020}}{\mathrm{M_\odot\,yr}^{-1}} \right) - 6 \log \left( \frac{T_\mathrm{eff}}{\text{K}} \right) ,
    \label{eq:temp_correction_sander+23}
\end{equation}
for $T_\mathrm{eff} > T_\mathrm{min}$. We assume a minimum temperature $T_\mathrm{min} = 1 \times 10^{5}$\,K \citep{Sander:2023aap}. For $T_\mathrm{eff} \leq T_\mathrm{min}$, the unmodified $\dot{M}_\mathrm{SV2020}$ is used.  

{Inevitably, we must make a number of simplifications to detailed models in order to adapt them to the limited information available in the evolutionary tracks used for rapid population synthesis.  For example, we take $T_\mathrm{eff}$ to be the effective temperature predicted by the rapid stellar evolution models \citep[][]{Hurley:2000MNRASSSE,COMPASTeam:2021tbl}, while \citet{Sander:2020MNRAS} use the temperature at the Rosseland continuum optical depth of 20 as a proxy for the temperature at the critical point.  Comparing our simplified prescriptions to the detailed models of \citet{Sander:2020MNRAS,Sander:2023aap} suggests that we match the detailed mass loss rate predictions for WR stars with a mass of 30 M$_\odot$, underpredict the mass loss rate for more massive stars (by $\sim 0.2$ dex at 40 M$_\odot$) and overpredict the mass loss rate for less massive stars (by $\sim 0.4$ dex at 20 M$_\odot$). Of course, even detailed models are incomplete.  For example, \citet{Sander:2020MNRAS} do not account for further radius dependencies \citep[e.g.,][]{Sander:2023aap} and dynamic inflation effects observed in multi-dimensional atmosphere models \citep[e.g.,][]{Moens:2022,Moens:2025}. Currently, there are no mass loss formulae consistently describing the mass loss of the full variety of observed WR and He stars, and we are thus limited to the available descriptions.}

The mass-loss rates given by \citet{Sander:2020MNRAS,Sander:2023aap} are applicable to massive, high-luminosity WR stars, and predict no mass-loss below a cutoff luminosity. 
For lower-mass (and thus lower-luminosity) helium stars, we opt to instead use the mass-loss rates from \citet{Vink:2017aap}
\begin{equation}
    \dot{M}_\mathrm{Vink2017} = -13.3 + 1.36 \log (L/\text{L}_\odot) + 0.61 \log (Z / \text{Z}_\odot) .
    \label{eq:MdotVink2017}
\end{equation}
following \citet{Woosley:2020ApJ}. 
Recently, a small sample of stripped helium stars has been unveiled \citep[e.g.,][]{DroutGotberg:2023Science,Gotberg:2023ApJ,Ramachandran:2024A&A}. 
\citet{Ramachandran:2024A&A} find that the \citet{Vink:2017aap} prescription provides an adequate estimate of the mass-loss rates of low-mass stripped stars, whilst \citet{Gotberg:2023ApJ} find that the mass-loss rates of their sample are lower than those predicted by \citet{Vink:2017aap}, suggesting that our model may overestimate the mass-loss rates for these stars. Future mass-loss prescriptions calibrated to the growing population of observed stripped stars can be implemented in COMPAS. 

The final mass-loss rate for helium stars is then determined as
\begin{equation}
    \dot{M}_\mathrm{WR} = \mathrm{max}(\dot{M}_\mathrm{Vink2017}, \dot{M}_\mathrm{SV2023})
    \label{eq:mdot_WR}
\end{equation}

Finally, as an alternative, we also implement the empirical WR mass-loss rate prescription from \citet{Shenar:2019aap}, based on a combined sample of analyzed WN stars from the Milky Way \citep{Hamann:2019A&A}, M31 \citep{Sander:2014A&A}, the SMC \citep{Hainich:2014A&A,Shenar:2016A&A}, and the LMC \citep{Hainich:2015A&A,Shenar:2019aap}.

In COMPAS, we include the option to multiply WR mass-loss rates by an arbitrary factor $f_\mathrm{WR}$ \citep[cf.][]{Barrett:2018MNRAS}. By default, we assume $f_\mathrm{WR} = 1$ for all plots in this paper. We compare the mass-loss prescriptions from \citet{Belczynski:2010ApJ}, \citet{Vink:2017aap}, \citet{Shenar:2019aap} and \citet{Sander:2020MNRAS} in Figure~\ref{fig:WR_mass_loss_rates}.  

\section{Population synthesis results}
\label{sec:results}

In this section we use the binary population synthesis code COMPAS to examine the impact of the updated mass-loss prescriptions for massive stars described in the previous section.

\subsection{Maximum black hole mass}
\label{subsec:MBH_max}

We begin by investigating the impact of our updated mass-loss prescriptions on the maximum black hole mass that can be formed by a single star.  
We evolve grids of single stars in the mass range 5---150\,M$_\odot$ and in the metallicity range $0.0001 \leq Z \leq 0.03$ \citep[][]{Hurley:2000MNRASSSE}. Unless otherwise stated, we use the default COMPAS settings, and only vary the mass-loss prescriptions. 
We calculate remnant masses and kicks using the remnant prescription from \citet{Mandel:2020qwb}.

We present results for three different sets of mass-loss prescriptions. These are the previous defaults $\textsc{$\textsc{$\textsc{Belczynski2010}$}$}$, our new defaults ($\textsc{Merritt2025}$), and a choice with higher mass loss in most evolutionary phases ($\textsc{Pessimistic}$). 
The choices of mass-loss prescription for each evolutionary phase (OB, VMS, RSG, WR) are given in Table~\ref{table:prescriptions}. 

\begin{figure}
    \centering
    \includegraphics[width=\columnwidth]{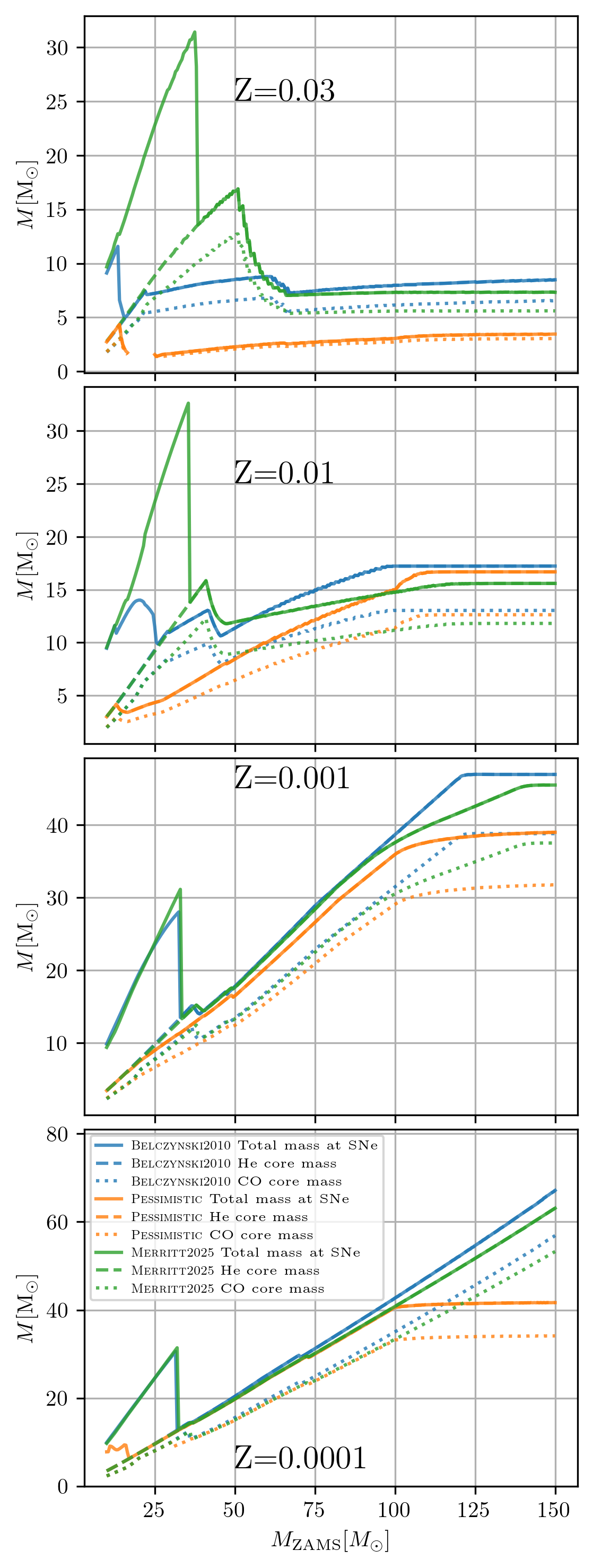}
    \caption{Initial-final total and core mass relation for single massive stars. The top panel shows the relation for $Z=0.03$ and the bottom panel for $Z=0.0001$, the highest and lowest metallicities modeled in COMPAS, respectively. }

    \label{fig:initial_final_mass_relation}
\end{figure}

In Figure \ref{fig:initial_final_mass_relation} we show the initial-final mass relation at $Z = [0.03, 0.01, 0.001, 0.0001]$.
 The `saw-tooth' feature at lower masses is due to stars experiencing SNe during core helium burning with an intact envelope, although $\textsc{Pessimistic}$ stars with increased RSG winds lose most of their hydrogen envelope even for relatively low ZAMS masses. With our updated RSG mass-loss rates ($\textsc{Merritt2025}$, using the \citealp{Decin:2024AA} prescription for RSGs), stars with $10\,\text{M}_\odot <  M_\mathrm{ZAMS} \lesssim 35\,\text{M}_\odot$ do not lose their entire hydrogen envelopes. If a significant fraction of the hydrogen envelope can fall back, this may result in a relatively high mass BH. Beyond this, the drop at ZAMS masses of $\sim 40 \,\rm M_\odot$ is due to envelope ejection/self-stripping, which occurs at an abrupt $Z$-dependent threshold in $M_\mathrm{ZAMS}$ driven primarily by RSG winds. Note that this feature is present with the $\textsc{Merritt2025}$ winds and to a lesser extent with $\textsc{$\textsc{Belczynski2010}$}$, but not with the $\textsc{Pessimistic}$ winds (essentially all BH progenitors self-strip). The plateaus around $\rm 100 \,\text{M}_\odot$ are due in part to the onset of LBV winds at the HD limit, and to VMS winds above 100$\,\rm M_\odot$.

\begin{figure*}
    \centering
    \includegraphics[width=\textwidth]{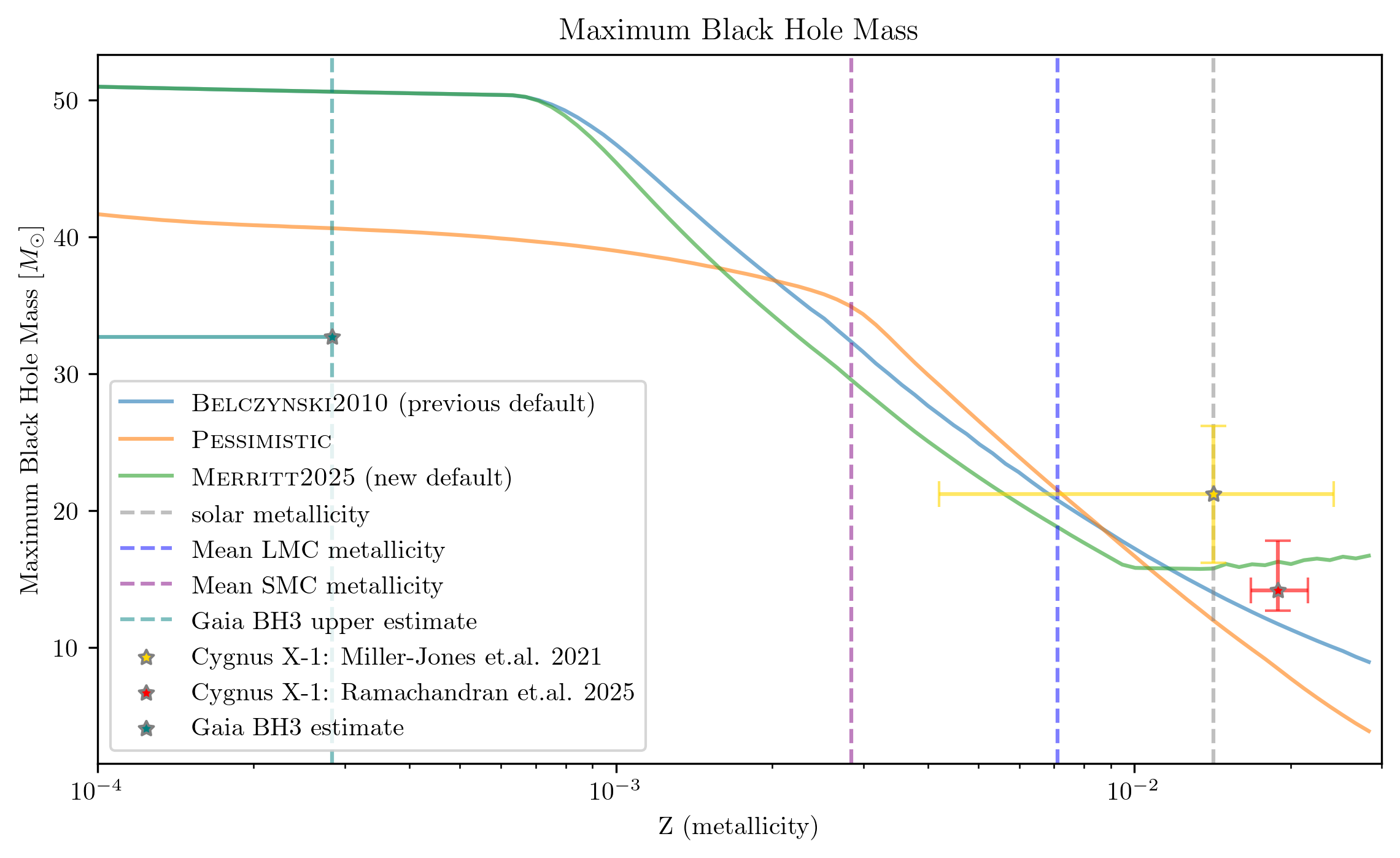}
    \caption{Maximum black hole mass formed from a single star as a function of metallicity.
    The blue line shows the result using the previous wind prescription in COMPAS \citep[][]{Belczynski:2010ApJ}.
    The green line shows the updated result using the new default combinations of wind prescriptions implemented in this work.
    The orange line shows a pessimistic choice of mass-loss prescriptions that lead to increased mass loss and typically lower black hole masses. 
All models use the \citet{Mandel:2020qwb} remnant mass prescription. 
    The maximum black hole mass at low metallicity ($Z \lesssim 10^{-3}$) for the $\textsc{Belczynski2010}$ and \textsc{Merritt2025} models is set by pair instability supernovae, implemented according to \citet{hendriks2023pulsationalpairinstabilitysupernovaegravitationalwave}.
    The yellow and red crosses and stars show the mass estimates for the black hole in Cyg X-1 from \citet{Miller-Jones:2021plh} and \citet{Ramachandran:2025}, respectively, while the leftward teal line shows the mass and metallicity for Gaia BH3 \citep[][]{GaiaBH32024}. LMC and SMC metallicity are assumed to be 0.5 and 0.2 of solar, respectively  \citep[e.g.,][]{Xshoot}.}
\label{fig:max_BH_mass_Z}
\end{figure*}

We show the maximum black hole mass formed as a function of metallicity in Figure~\ref{fig:max_BH_mass_Z}.
At low metallicity ($Z \lesssim 10^{-3}$), the maximum black hole mass is determined by the onset of the electron-positron pair-instability \citep[][]{Fowler:1964ApJS,Rakavy:1967ApJ}, leading to a mass gap above a black hole mass of around $45$\,M$_\odot$. 
For stars experiencing pulsational pair-instability supernovae \citep[PPISN][]{Belczynski:2016A&A,Woosley:2017ApJ,Farmer:2019ApJ,Marchant:2019ApJ,Stevenson:2019ApJ,hendriks2023pulsationalpairinstabilitysupernovaegravitationalwave, 2024arXiv240716113R}, we employ the \citet{hendriks2023pulsationalpairinstabilitysupernovaegravitationalwave} prescription, based on the detailed stellar models from \cite{Farmer:2019ApJ}, \citep[see also][]{2022RNAAS...6...25R}. 
This model employs a flexible parameter for the onset of PPISN, where the default is above a carbon-oxygen core mass of $34.8 \,\text{M}_\odot$.

At higher metallicity ($Z \gtrsim 10^{-2}$), some of the most massive black holes form via RSG progenitors that avoid losing their hydrogen envelopes through stellar winds.
The maximum black hole mass in this metallicity range is $\sim 32$\,M$_\odot$, using the \cite{Fryer_2012} remnant mass prescription (which allows stars to retain their hydrogen envelopes upon collapse), and $\sim18$\,M$_\odot$, using the \cite{Mandel:2020qwb} remnant mass prescription, which assumes that the hydrogen envelope is expelled via the \citet{Nadezhin:1980Ap&SS} mechanism \citep[][]{Lovegrove:2013ApJ,Fernandez:2018MNRAS}. Several authors have argued that it may be possible to form more massive black holes if the hydrogen envelope is retained and falls back onto the black hole \citep[e.g.,][]{Costa:2021MNRAS,Farrell:2021MNRAS,Vink:2021MNRASmaxBH}.  On the other hand, the hydrogen envelope is expected to be stripped by mass transfer prior to BH formation in compact binaries, such as those leading to the formation of merging BHs and X-ray binary systems.

\subsection{Forming the most massive Galactic stellar mass BHs: Cyg X-1 and Gaia BH3}
\label{subsec:CygX1}

Until recently, the most massive stellar-mass black hole known in the Milky Way was in the high-mass X-ray binary Cygnus X-1, with a mass of $21 
\pm 2$\,M$_\odot$ \citep{Miller-Jones:2021plh}. A recent effort from \citet{Ramachandran:2025} reduces the estimate to $17.5^{+2}_{-1} 
M_\odot$, and further to $14.2^{+3.6}_{-1.5}M_\odot$ (the estimate appearing in Figure~\ref{fig:max_BH_mass_Z}) accounting for the uncertainty in the orbital inclination, and refines the estimate of metallicity to $1.33^{+0.19}_{-0.15}Z_{\odot}$. Their models predict that continued evolution may lead to a BBH merger in $\sim5Gyr$. Because the system appears to have formed at super-solar metallicity, this is a unique test of our models.
The black holes in high-mass X-ray binaries like Cyg X-1 are thought to form from primary stars stripped of their hydrogen envelope through mass transfer \citep[e.g.][]{Valsecchi:2010Natur,Neijssel:2021imj,2022ApJ...938L..19G}.
Using COMPAS, \citet{Neijssel:2021imj} and \citet{Romero-Shaw:2023MNRAS} found that they required a reduction of WR mass-loss rates at solar metallicity by a factor of 3--10 in order to account for the formation of a black hole with a mass similar to Cyg X-1. 
This is essentially because the most massive helium core mass formed at solar metallicity is only around 25$\,\text{M}_\odot$, leaving little room for mass loss.
With our updated mass-loss prescription, we find that we still cannot form a $20M_{\odot}$ black hole at solar metallicity or higher without artificially reducing the WR mass-loss rates. Our updated models can match the match the lower mass estimate ($14.2M_{\odot}$) of \citet{Ramachandran:2025}, forming up to $17M_{\odot}$ black holes at super-solar metallicity. 

\citet{Higgins:2021MNRAS} evolve detailed models of helium stars using WR mass-loss rates from \citet{Sander:2020MNRAS}. 
They find that a helium star with $M_\mathrm{He,i} \gtrsim~40$\,M$_\odot$ can produce a helium star with a final mass of $M_\mathrm{He,f} \sim 20$\,M$_\odot$, capable of producing a black hole as massive as Cyg X-1. 
However, they do not address the question of forming such a massive helium star at solar metallicity. 

One potential pathway for forming high helium core masses at solar metallicity is to form Cyg X-1 through chemically homogeneous evolution \citep[CHE;][]{deMink:2009A&A,Qin:2019ApJL}. 
In this scenario, rapid rotation due to tidal locking induces efficient mixing within the primary star, allowing it to fuse nearly all of its hydrogen into helium, thus producing a more massive helium star than classical evolution would allow.

Perhaps the most obvious resolution of the tension between the predicted and observed black hole masses may be due to shortcomings of stellar and binary evolution models in COMPAS.  In particular, the use of \citet{Hurley:2002MNRASBSE} models for the evolution of mass-losing main-sequence stars leads to an artificial decrease in the final core mass (see discussion in \citealt{Romero-Shaw:2023MNRAS}).  
Mass loss during the main sequence affects both the radial evolution of the mass-losing main sequence star, allowing it to remain compact and blue \citep[][]{Bavera:2023NatAs} and thus impacting subsequent mass transfer, and the evolution of the convective core, with detailed models allowing more massive helium-rich cores to form than in COMPAS models \citep[][]{shikauchi2024evolutionconvectivecoremass}.
The implications of allowing mass-losing stars to retain a greater fraction of their core mass \citep{shikauchi2024evolutionconvectivecoremass} for the formation of Cygnus X-1 will be discussed in detail elsewhere (Brcek et al., in prep.).

Assumptions about convective mixing and overshooting also impact predictions for the final core masses of stars.  COMPAS uses the stellar models from \citet{Pols:1998MNRAS}.   Examining five sets of commonly used stellar models at solar metallicity, \citet[][]{Agrawal:2022MNRAS} show that the maximum core mass that can be formed at solar metallicity varies in the range 20--40\,M$_\odot$ under different model assumptions. 
Other studies \citep[e.g.,][]{Zapartas:2021A&A,Martinet:2023A&A,Vink:2024A&A} also find a broad range of maximum black hole masses.
\citet{Bavera:2023NatAs} found in their detailed binary simulations that not only can black holes up to 30\,M$_\odot$ form at solar metallicity, they can be members of merging binary black holes.

\citet{Romagnolo:2024ApJL} recently implemented a series of updated mass-loss prescriptions for massive stars in a \textit{detailed} stellar evolution code to examine the maximum black hole mass that can be formed at solar metallicity from a slowly rotating, single star. 
They use the mass-loss prescription from \citet[][]{Bestenlehner:2020MNRAS} for VMS. 
For their choice of mass-loss prescriptions and input physics, they are unable to produce black holes with masses $\gtrsim 30$\,M$_\odot$, unlike the results of \citet{Bavera:2023NatAs} and \citet{Martinet:2023A&A}. 

The recent discovery of Gaia BH3 \citep{GaiaBH32024}, a $33$\,M$_\odot$ black hole in a 11.6\,yr orbit with a low-mass companion star, has revealed the existence of a population of even more massive stellar mass black holes in the Milky Way. 
Gaia BH3 may have formed through isolated binary evolution from an initially wide, eccentric binary \citep[][]{El-Badry:2024GaiaBH3,Iorio:2024GaiaBH3}.
Alternatively, Gaia BH3 may have formed dynamically in a star cluster \citep[][]{MarinPina:2024GaiaBH3}, as hinted at by the association with the disrupted stream ED-2 \citep[][]{GaiaBH32024,Balbinot:2024GaiaBH3}.
While the birth metallicity of Gaia BH3, at [Fe/H] $\sim -2.5$ \citep[][]{Balbinot:2024GaiaBH3}, which would correspond to $Z<10^{-4}$, is outside of the range that can be modeled using COMPAS \citep[][]{Pols:1998MNRAS,Hurley:2000MNRASSSE,COMPASTeam:2021tbl}, this BH likely did not form from the current Population I stars. We see from Figure~\ref{fig:max_BH_mass_Z} that with our updated mass-loss prescriptions, black holes with masses $\gtrsim 30$\,M$_\odot$ can form at metallicities $Z \lesssim 3 \times 10^{-3}$, consistent with the identification of the companion star as a very metal poor star \citep[][]{GaiaBH32024}.

\subsection{Impact on double compact objects}
\label{subsec:DCO_impact}

\begin{figure*}
    \centering
    \includegraphics[width=\textwidth]{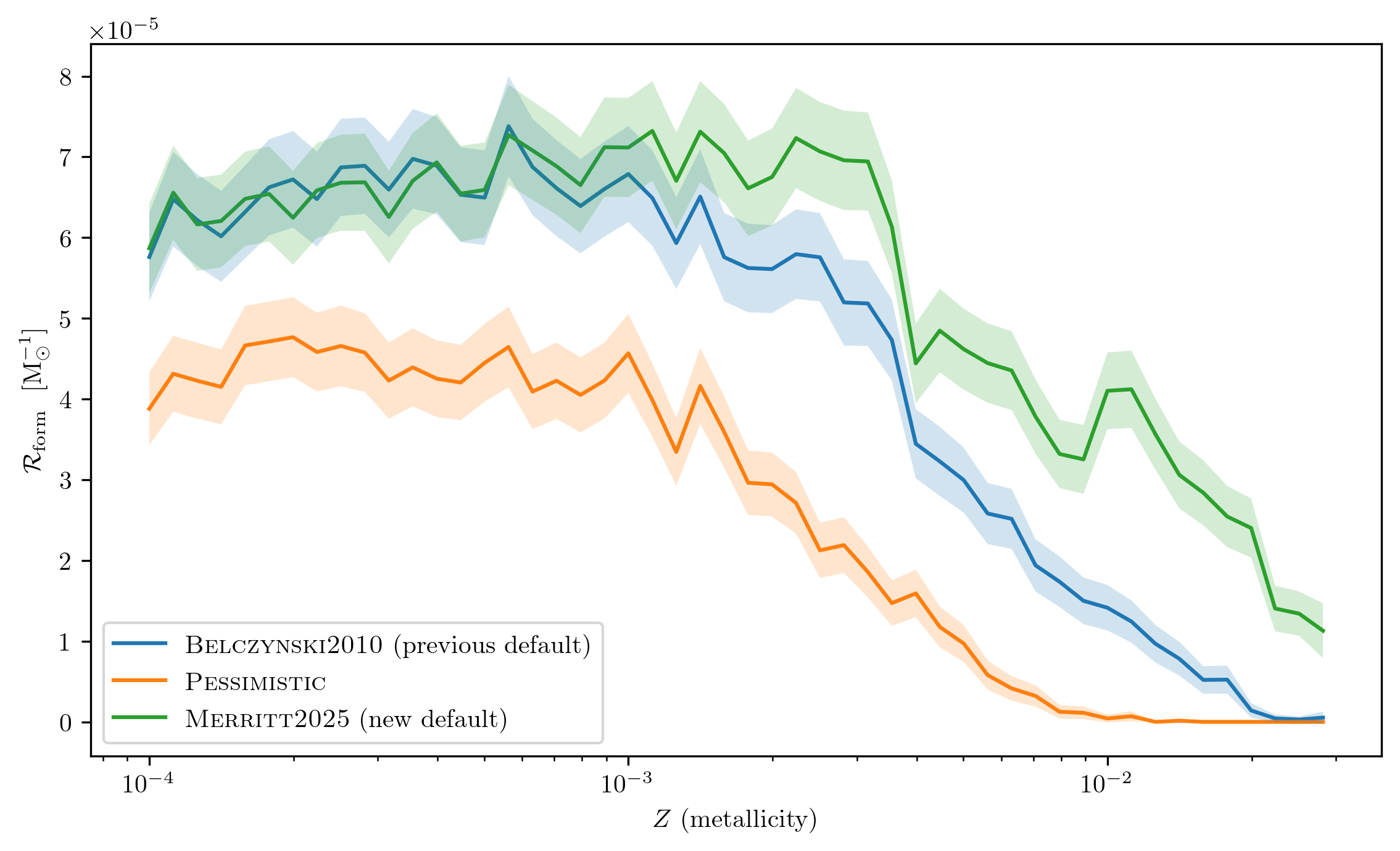}
    \caption{Formation rates of merging BBHs per star forming mass as a function of metallicity. The different colored lines denote our different sets of mass-loss prescriptions (see Table~\ref{table:prescriptions}). The shaded region around each line indicates the sampling uncertainty at $\pm2\sigma$. The blue line shows the previous default model \citep[$\textsc{Belczynski2010}$: ][]{Belczynski:2010ApJ,COMPASTeam:2021tbl}, the orange line shows the \textsc{Pessimistic} model and the green line shows our new default model, $\textsc{Merritt2025}$.  
    }
    \label{fig:BBH_yields_vs_Z}
\end{figure*}

A large sample of double compact object (DCO) mergers have now been observed in gravitational waves \citep[][]{Abbott:2023PhRvXGWTC3,Abac:2025GWTC4}, including binary neutron star \citep[BNS;][]{Abbott:2017PhRvL}, neutron star-black hole \citep[NSBH;][]{Abbott:2021ApJL} and binary black hole \citep[BBH;][]{Abbott:2016PhRvL} mergers.
DCOs may form via the evolution of massive stellar binaries \citep[e.g.,][]{1973NInfo..27...70T, 1973A&A....25..387V, 2021hgwa.bookE..16M, 2022PhR...955....1M}.

In this section, we perform population synthesis using COMPAS to examine the impact of our updated mass-loss prescriptions on the formation of merging DCOs. 
Unlike \citet{Riley:2020btf}, we assume that binaries experiencing Roche lobe overflow at birth lead to stellar mergers, because including these leads to an overproduction of BBHs through CHE inconsistent with observation \citep[][]{Stevenson:2022MNRAS}.
We assume the \citet{Mandel:2020qwb} remnant mass prescription and kick distribution. 
We calculate the yields of DCOs that merge within the age of the universe formed at a given metallicity per unit of star formation. 
In Figure~\ref{fig:BBH_yields_vs_Z} we show the yield of merging BBHs as a function of metallicity for each of our sets of mass-loss prescriptions. 

Yields of double compact objects are highly sensitive to the treatment of mass loss \citep[e.g.,][]{Broekgaarden:2022MNRAS,Wagg:2022ApJ}. 
We find that the treatments of RSG and WR winds are most consequential, especially on the formation of BBHs. 
WR winds have a dominant effect on the production of high metallicity systems, because high-mass stars are almost universally stripped above solar metallicity. 

At low metallicity we find that RSG winds have a dominant effect on yields. The $\textsc{Pessimistic}$ \citep{VinkSabhahit:2023aap} RSG wind prescription leads to a reduction in merging BBH yields by a factor of ${\sim}2$. At these metallicities and below a ZAMS mass of 40$\,\text{M}_\odot$ (frequently the secondary for BBH formation), whether the star self-strips depends on the choice of prescription, as mass-loss rates vary over 3-5 orders of magnitude. 
By this time the core mass is sufficient for black hole formation in any case, but the ejection of the envelope often leads to orbital widening that prevents the system from merging in a Hubble time. 

Compared to BBHs, the yields of BNSs and NSBHs, which generally come from less massive progenitors with lower wind mass loss,  are more robust to changes in the mass-loss prescription, being consistent with sampling uncertainties at most metallicities. Our findings match those of \citet{vanson2024windschangebinaryblack}.

\begin{figure*}\centering
    \includegraphics[width=\textwidth]{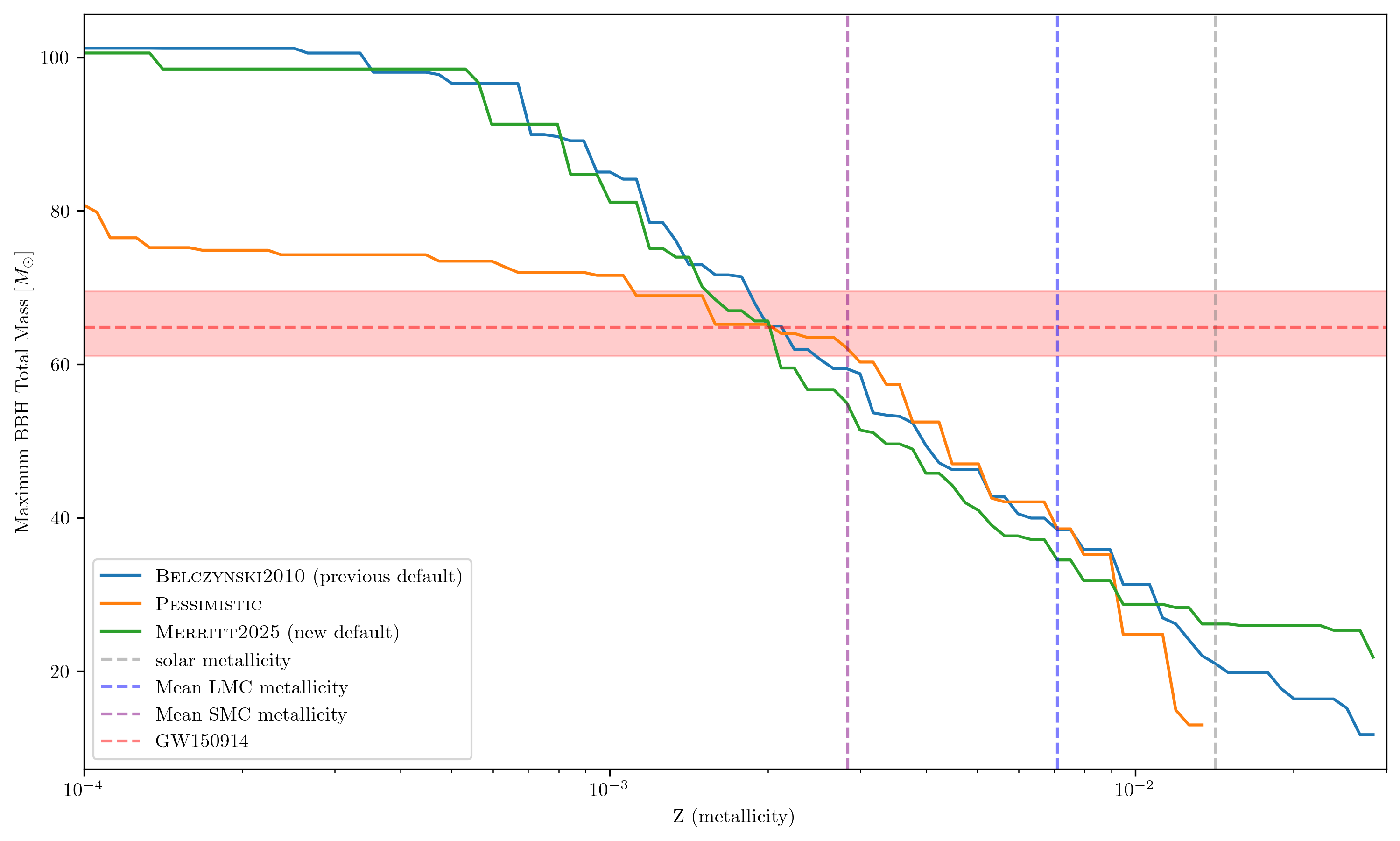}
    \caption{Maximum total mass of merging BBHs as a function of metallicity, as distinct from final mass which would account for energy lost in GWs.
    The vertical dashed lines denote solar metallicity, and that of the Large and Small Magellanic Clouds.
    The horizontal dashed red line denotes the total mass of GW150914, with the median sample, and 90\% credible intervals filled \citep{Abbott:2016PhRvL}, illustrating that such high-mass BBHs are expected to form only in low-metallicity environments in these models. Note that under the $\textsc{Pessimistic}$ (orange) model, merging BBH of any mass do not form at or above solar metallicity. }
    \label{fig:max-Mtot-func-Z}
\end{figure*}

\begin{figure*}
    \centering
    \includegraphics[width=\textwidth]{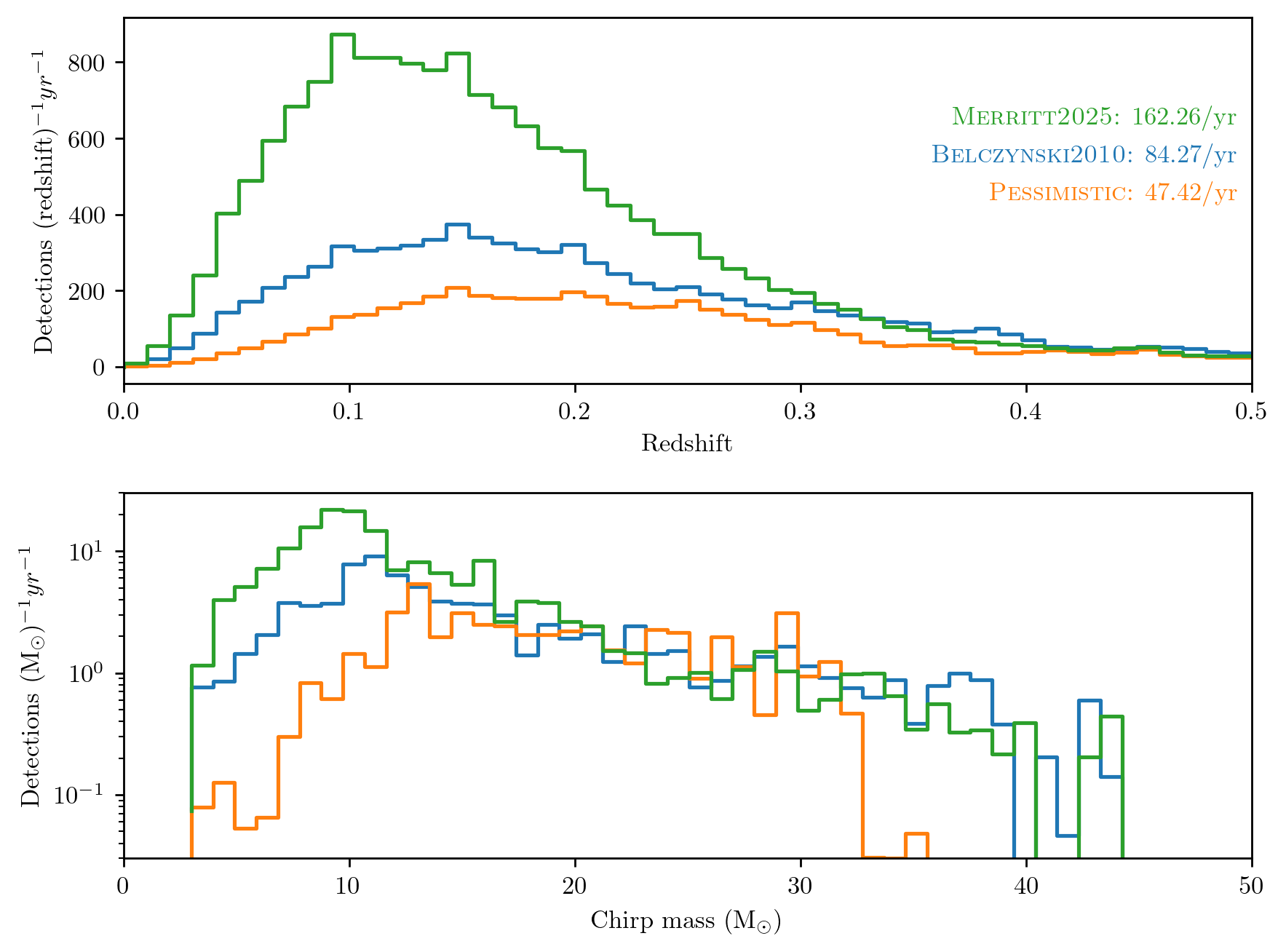}
    \caption{Projected detections of BBH binned by chirp mass and redshift, assuming sensitivity of the third observing run (O3) of LVK, with a detection threshold SNR of 8. Updating winds creates an excess near a chirp mass of 10\,M$_\odot$, and slightly lowers the density at 25--45\,M$_\odot$. Blue=$\textsc{Belczynski2010}$, Orange=$\textsc{Pessimistic}$, Green=$\textsc{Merritt2025}$.
    }
    \label{fig:BBH-Chirp-Mass}
\end{figure*}

In Figure~\ref{fig:max-Mtot-func-Z} we show the maximum total mass of merging BBHs formed as a function of metallicity \citep[see also][]{Belczynski:2016Nature}. 
Similar to the single star case (Figure~\ref{fig:max_BH_mass_Z}), we find a large difference at low-metallicity ($Z = 10^{-4}$) when using the $\textsc{Pessimistic}$ model, with a maximum total mass of merging BBHs of around $75$\,M$_\odot$ compared to $90$\,M$_\odot$ in $\textsc{Belczynski2010}$ and $\textsc{Merritt2025}$. 
This means that massive merging BBHs such as GW150914 would need to be formed at $Z \lesssim 10^{-3}$ with the $\textsc{Merritt2025}$ and $\textsc{Belczynski2010}$ prescriptions or at even lower metallicities with the $\textsc{Pessimistic}$ winds prescription \citep[][]{Abbott:2016ApJL,Belczynski:2016Nature}.
At solar metallicity ($Z = 0.0142$), we find that our new $\textsc{Merritt2025}$ model can produce merging BBHs with total mass up to $26$\,M$_\odot$, compared to $21$\,M$_\odot$ in $\textsc{Belczynski2010}$, while no merging BBHs are produced under the $\textsc{Pessimistic}$ model. 
At intermediate metallicities typical for the SMC and LMC, the three models predict similar maximum total BBH masses. 

In Figure~\ref{fig:BBH-Chirp-Mass} we show predictions for the observed chirp mass and redshift distributions of merging BBHs. 
We perform cosmic integration using the built in tools in COMPAS  \citep{Barrett:2018MNRAS,Neijssel_2019,COMPASTeam:2021tbl}. 
We assume a gravitational-wave network with the approximate sensitivity of LIGO during the O3 run. 
In the new $\textsc{Merritt2025}$ model, the chirp mass distribution  has an increased pileup in the 10\,M$_\odot$ peak and a decrease at higher masses. 
We also find that the total predicted BBH detection rate increases by a factor of ${\sim}2$ in the $\textsc{Merritt2025}$ model compared to the $\textsc{Belczynski2010}$ model. 
The $\textsc{Pessimistic}$ model decreases the BBH detection rate by a factor of ${\sim}2$ compared to the $\textsc{Belczynski2010}$ model.
For BNS and NSBH mergers, the chirp mass distribution remains approximately the same, within sampling error, among the tested wind variations. 

\section{Discussion and conclusions}
\label{sec:discussion}

Massive stars can lose a significant fraction of their mass through stellar winds in various evolutionary phases, impacting the masses of the compact objects they leave behind. 
Mass-loss prescriptions  are hence an important ingredient in any stellar evolution or population synthesis code.
There has been significant progress recently in both theoretical predictions and empirical measurements of the mass-loss rates for massive stars. However, large uncertainties remain.  Making use of this progress, and taking advantage of the ability of rapid population synthesis tools to explore a broad range of scenarios, we have updated the mass-loss rate prescriptions for several phases of massive stellar evolution (MS, VMS, RSG and WR) in the rapid binary population synthesis code COMPAS \citep[][]{Stevenson:2017tfq,COMPASTeam:2021tbl}. 
In total, we have implemented 15 new mass-loss prescriptions.

Our main caveats and conclusions are:
\begin{itemize}
    \item [--] Mass loss rates for massive stars (with $10 < M_\mathrm{ZAMS} / $\,M$_\odot\, < 100$) on the MS are reasonably well constrained, and even the factor of $2$--$3$ uncertainty in these rates \citep[][]{Vink:2001aap,Krticka:2018aap,Bjorklund:2023aap} is likely to be subdominant to the impact of uncertainties in mass-loss rates during later evolutionary phases.
    
    \item [--] For massive RSGs, it appears that there is still a large amount of variation in the determination of mass-loss rates \citep[][]{Wen:2024arXiv}. 
    In the case of high mass-loss rates \citep[e.g.,][]{Yang:2023arXiv,VinkSabhahit:2023aap}, this leads to RSGs being able to lose their hydrogen envelopes, potentially providing a solution to the missing red supergiant problem \citep[][]{Smartt_2009}. Alternatively, low mass-loss rates \citep[][]{Beasor:2020MNRAS,Beasor:2023MNRAS,Decin:2024AA} imply that massive RSGs are unable to lose their hydrogen envelopes through steady-state winds alone, although pulsations/eruptive mass loss may still provide an alternative mechanism to remove the envelope. 

    \item [--] Recently, \citet{Wen:2024arXiv} and \citet{Antoniadis:2024arXiv} have presented empirical studies of the mass-loss rates of large samples of RSGs in the Large Magellanic Cloud (LMC), similar to that of \citet{Yang:2023arXiv} for the SMC. 
    These studies identify a turning point in the trend of mass-loss rates for stars with luminosity at around $\log L/\text{L}_\odot = 4.4$.
    Below this luminosity, they find RSG mass-loss rates at least an order of magnitude lower than those of \citet{Yang:2023arXiv} and other contemporary RSG prescriptions.
    This finding may indicate a systematic uncertainty due to differences in modeling approaches between \citet{Wen:2024arXiv}, \citet{Antoniadis:2024arXiv} and \citet{Yang:2023arXiv}. 
    These recent studies indicate that while empirical samples of RSGs are becoming larger and better studied, we still do not have a complete understanding of RSG mass-loss and its scaling with metallicity.

    \item [--] Empirical mass-loss rate prescriptions are predominantly calibrated only at the relatively high metallicities of the Milky Way, the Magellanic Clouds \citep[e.g.,][]{Yang:2023arXiv,Wen:2024arXiv} and other nearby galaxies \citep[e.g.,][]{Wang:2021ApJ}.  These prescriptions must then be extrapolated to the much lower metallicities ($Z \lesssim 10^{-3}$) relevant for binary black hole formation. Recently, \citet{Pauli:2025A&A} presented a new mass-loss description derived from empirical results in the Milky Way, LMC, and SMC. They could show that to first order the winds from main sequence OB stars do not differ from those of binary-stripped stars when expressing the mass-loss rate as a function of $\Gamma_\text{e}$ and initial $Z$. However, there is a considerable scatter as further parameters are not taken into account. In principle, the description from \citet{Pauli:2025A&A} would provide an alternative to both the ``OB'' and ``WR'' treatments ($\lesssim\,100\,$kK), but this needs to be explored in a separate study.

    \item[--] With our updated winds prescription (including reduced WR mass-loss rates), we are unable to produce black holes at solar metallicity more massive than ${\sim}20$\,M$_\odot$. However, the details of the stellar evolution models used may be able to help explain this discrepancy, including models of stellar response to mass loss \citep[][]{Agrawal:2020MNRAS,Agrawal:2022MNRAS, shikauchi2024evolutionconvectivecoremass}.

    \item[--] With our updated winds prescription, massive stellar-mass black holes such as Gaia BH3 \citep[][]{GaiaBH32024} can form at metallicities $Z \lesssim 3 \times 10^{-3}$.

    \item[--] The yield of merging BBHs is sensitive to uncertainties in mass-loss rates, especially in WR and RSG phases. The yields of merging BNSs and NSBHs are much less sensitive to uncertainties in mass-loss rates, giving similar predictions to \citet{vanson2024windschangebinaryblack}.

    \item[--] The code modifications described in this paper were completed in 2023.  They already mark a significant improvement over the standard set still used in many other population synthesis codes (and even some detailed modeling efforts using MESA). We envisage ongoing improvements in the prescriptions to track ongoing observational and theoretical advances in mass-loss models.
    
\end{itemize}

\vspace{15pt}
The authors acknowledge support from the Australian Research Council (ARC) Centre of Excellence for Gravitational Wave Discovery (OzGrav), through project number CE170100004 and CE230100016.
Simon Stevenson is a recipient of an ARC Discovery Early Career Research Award (DE220100241).
AACS is funded by the Deutsche Forschungsgemeinschaft (DFG, German Research Foundation) in the form of an Emmy Noether Research Group -- Project-ID 445674056 (SA4064/1-1, PI Sander). AACS further acknowledges support from the Federal Ministry of Education and Research (BMBF) and the Baden-Württemberg Ministry of Science as part of the Excellence Strategy of the German Federal and State Governments.  The work of AACS and IM was performed in part at the Aspen Center for Physics, which is supported by National Science Foundation grant PHY-2210452.
BF acknowledges support from the National Science Foundation under grant PHY-2146528, OzGrav through their international visitor program, the Simons Foundation through their sabbatical visitor program, and CERN through their Scientific Associate program.
TW acknowledges support from NASA ATP grant 80NSSC24K0768.

\vspace{5mm}

\software{
COMPAS v03.14.00 \citep[][]{COMPASTeam:2021tbl,Compas:2022JOSS,COMPAS:2025}. 
Our input files and analysis are available on Zenodo: \dataset[DOI: 10.5281/zenodo.16331862]{https://doi.org/10.5281/zenodo.16331862}.}

\bibliography{bib}{}
\bibliographystyle{aasjournal}

\end{document}